\documentclass[aps,twocolumn,pra,superscript,floatfix,superscriptaddress,showpacs]{revtex4-1}

\usepackage{graphicx}% Include figure files
\usepackage{dcolumn}% Align table columns on decimal point
\usepackage{bm}% bold math
\usepackage{color,soul}

\usepackage{amsmath}

\usepackage[colorlinks=true,%
            linkcolor=blue,%
            urlcolor=blue,%
            citecolor=blue,%
            filecolor=blue,%
            bookmarksopen=true,%
            pdfauthor={AGarrido},%
            pdftitle={QDinterfometerMZM},%
            pdfsubject={QDinterfometerMZM_Manuscript},%
            pdfpagemode=UseOutlines]{hyperref}

\graphicspath{{Images/}} %Images folder

\definecolor{applegreen}{rgb}{0.55, 0.71, 0.0}
\definecolor{antiquefuchsia}{rgb}{0.57, 0.36, 0.51}
\definecolor{amethyst}{rgb}{0.6, 0.4, 0.8}

\begin{document}

%\preprint{AAPM/123-QED}

\title{Bound states in the continuum and Majorana zero modes in a double quantum dot interferometer: Ghost-Fano Majorana effect\footnote{Error!}}

\author{A. P. Garrido}
\email{alejandro.garridoh@usm.cl}
\affiliation{ 
Departamento de F\'isica, Universidad T\'ecnica Federico Santa Mar\'ia, Av. Espa\~na 1680, Casilla 110V, Valparaiso, Chile.%\\This line break forced with \textbackslash\textbackslash
}%

\author{D. Zambrano}
 \affiliation{ 
Departamento de F\'isica, Universidad T\'ecnica Federico Santa Mar\'ia, Av. Espa\~na 1680, Casilla 110V, Valparaiso, Chile.%\\This line break forced with \textbackslash\textbackslash
}%
\author{J. P. Ramos-Andrade}
\affiliation{Departamento de F\'isica, Universidad de Antofagasta, Av. Angamos 601, Casilla 170, Antofagasta, Chile.}
\author{P. A. Orellana}%
 \affiliation{ 
Departamento de F\'isica, Universidad T\'ecnica Federico Santa Mar\'ia, Av. Espa\~na 1680, Casilla 110V, Valparaiso, Chile.%\\This line break forced with \textbackslash\textbackslash
}%

\date{\today}% It is always \today, today,
             %  but any date may be explicitly specified

\begin{abstract}
We investigate the transport properties through a nanostructure composed of parallel double quantum dots coupled to two normal contacts. Additionally, each quantum dot is connected to a topological superconducting nanowire, hosting Majorana zero modes at its ends. A magnetic flux threading across the area enclosed by the interferometer is considered. First, we investigate the physical quantities of the system employing Green's function formalism. We find that the emergence of bound states appears in symmetric configurations of topological superconducting nanowires, i.e., depending on their lengths and coupling energies to the quantum dots. Also, we find a transport suppression anomaly as a function of the magnetic flux in the same symmetric configurations mentioned above. Besides, we find that the magnetic flux controls both the projection of Majorana zero modes and of the bound states in the continuum into the density of states and the linear conductance, suggesting that only by switching this parameter can we manipulate both bound states.
\end{abstract}

%Use showkeys class option if keyword.
                              %display desired
\maketitle

%%%%%%%%%%%%%%%%%%%%%%%%%%%%%%%%%%%%%%%%%%%%%%%%%%%%%%%%%%%%%%
\section{\label{sec:intro}Introduction}
%%%%%%%%%%%%%%%%%%%%%%%%%%%%%%%%%%%%%%%%%%%%%%%%%%%%%%%%%%%%%%

In recent years, the study of topological superconductor nanowires (TSCNs) has received a great deal of attention in condensed matter physics due to their potential for technological applications in quantum computing \cite{kitaev2003fault, nayak2008non, pachos2012introduction, beenakker2013search, laflamme2014publisher, albrecht2016exponential}. In this context, the existence of exotic fermionic quasiparticles has been predicted as quasiparticles that would be their own anti-quasiparticles \cite{majorana1937teoria, wilczek2009majorana,franz2010race}, as the ones appearing localized in topological superconducting systems. Due to their resemblance with Majorana fermions they are called Majorana zero modes (MZMs).

MZMs satisfy non-Abelian statistics and they can be manipulated by braiding operations \cite{kraus2013majorana, alicea2011non}, making them exceptional candidates for quantum computation implementations \cite{kitaev2001unpaired,bravyi2002fermionic,kitaev2003fault,nayak2008non,leijnse2011quantum,pachos2012introduction,kraus2013majorana,albrecht2016exponential}. Among others, MZMs systems are predicted to be found at the ends of a TSCN, composed of a semiconductor-superconductor nanowire with strong spin-orbit interaction in the presence of a magnetic field. This system can be seen as a setup of a Kitaev chain \cite{kitaev2001unpaired, kitaev2003fault,moore2009next}, in which the coupling between the two MZMs located at the wire's opposite ends decays exponentially with the wire's length \cite{albrecht2016exponential}, allowing to build of a qubit which is topologically protected from decoherence by local perturbations \cite{wu2012tunneling,kitaev2001unpaired,kraus2013majorana,albrecht2016exponential,semenoff2006teleportation,tewari2008testable}. 

The first physical realization of this system was achieved by Mourik and collaborators, announcing zero-bias anomalies in the conductance as a signature of the MZMs presence \cite{mourik2012signatures}. Later, many systems have been proposed \cite{bolech2007observing, nilsson2008splitting, law2009majorana, fu2009josephson, flensberg2010tunneling, pikulin2012h, franz2013majorana, prada2012transport,rainis2013towards, cook2012stability, liu2013manipulating, stanescu2011majorana, lee2014spin, wimmer2011quantum}, and several experiments based on zero-bias anomalies in transport properties through source-drain leads have been performed \cite{deng2012anomalous,mourik2012signatures,das2012zero,lee2012zero,finck2013anomalous,churchill2013superconductor,zambrano2018bound,ramos2019fano}. But these anomalies are not always a reliable evidence of MZMs, leading to the necessity of devising custom-made experimental protocols that allow e.g. performing simultaneous tunneling and Coulomb blockade spectroscopy measurements within the same device, in order to rule out MZMs detection ambiguities \cite{valentini2022majorana}.

On the other hand, the so-called bound states in the continuum (BICs) do not decay even if their energy levels are within the range of the continuum states \cite{hsu2016bound}. The BICs, predicted by von Neumann and Wigner \cite{vonNeumann-Wigner}, have been receiving great interest in photonic systems. Moreover, due to the typical interference phenomena analogy between electronic and photonic systems, the inherent possibility of BICs presence in electronic systems arises \cite{hsu2016bound, ramos2014bound, grez2022bound}. In this context, the electronic transport through quantum dots (QDs) structures has been an active research field during the past decades \cite{van2002electron, hanson2007spins, holleitner2001coherent, holleitner2002probing, shangguan2001quantum, orellana2003transport}. QDs are nanostructures with quantized energy levels due to the confinement of electrons, so they are usually called artificial atoms \cite{van2002electron}. Additionally, electrons tunneling through QDs show a high coherence preservation, demonstrated in several phenomena such as the subtle Kondo effect in QD connected to leads \cite{das2012zero, hofstetter2001kondo,gorski2019spin}, the Aharonov-Bohm (AB) oscillations in closed interferometers \cite{chi2007fano, kubala2002flux}, and Fano resonances in systems with multiple channels \cite{fano1961effects, miroshnichenko2010fano, van2002electron, hanson2007spins, chi2007fano, hofstetter2001kondo, de2003ghost}, among others.

In the past years, a wide range of research has been done regarding the effects of quantum interference in several configurations of the components previously explained: parallel, series, and T-shaped. In the systems of hybridized QD-TSCN, there is usually more than one electron transport path, and quantum interference effects are an efficient way to detect the existence of the MZMs formed at the ends of the TSCN \cite{gong2014detection, chi2021quantum}. For instance, in a non-interacting QD-leads system with a side coupled TSCN, Liu and Baranger established a particular signature of the presence of MZMs, which is a half-integer conductance at zero-energy \cite{liu2011detecting}. Later, it was shown that this zero-bias anomaly is due to MZM leaking into the QD \cite{vernek2014subtle}, and it is robust against changes in the QD energy level, which was shortly after experimentally verified \cite{deng2016majorana}. Aditionally, in QD-MZMs systems, where the interplay between MZM-BICs can take place, a theoretical encryption setup based on BICs \cite{ricco2016decay} and Majorana fermion qubits readout technology \cite{guessi2017encrypting} have been proposed.

In previous work, a double QD (DQD) interferometer has shown an anomaly of suppressed transport and signatures of a flux-dependent level attraction, which can be manipulated by an applied magnetic flux and gate voltages \cite{kubala2002flux}. Moreover, when a TSCN is coupled to the DQD, the linear conductance shows MZM signatures at zero energy and then inducing the Fano effect \cite{chi2021quantum}.

Within this context, in the present work we study a system formed by a DQD structure coupled to two normal leads, while each QD is independently connected to a TSCN hosting MZMs at its ends. We focus on the linear conductance through the DQD, the QD's density of states, and the MZMs spectral functions, which are calculated employing the Green's functions (GFs) formalism. We focus in identifying signatures of quantum interference phenomena, MZMs leakage into the QDs-BICs, and the interplay between MZM and BIC, by direct control of the magnetic flux over all the bound states of our setup. Our results show that both MZMs and BICs appear in high-symmetry configurations, i.e., depending on the QD-MZM coupling strength and the length of the TSCN. Also, we find a transport suppression anomaly in the linear conductance as a function of the magnetic flux. This phenomenon appears for the same symmetric configurations mentioned above. We also find that both the MZMs leaking into the QDs and the BICs can be controlled by the magnetic flux, suggesting that this external parameter will suffice for manipulating the above states.

This paper is organized as follows: Section\ \ref{secII} presents the model and method used to obtain quantities of interest; Section\ \ref{secIII} shows the results and discussions, and the concluding remarks are presented in Section\ \ref{secIV}.

%%%%%%%%%%%%%%%%%%%%%%%%%%%%%%%%%%%%%%%%%%%%%%%%%%%%%%%%%%%%%%
\section{Model and method}\label{secII} 
%%%%%%%%%%%%%%%%%%%%%%%%%%%%%%%%%%%%%%%%%%%%%%%%%%%%%%%%%%%%%%

\begin{figure}
    \centering
    \includegraphics[width=26em,height=10em]{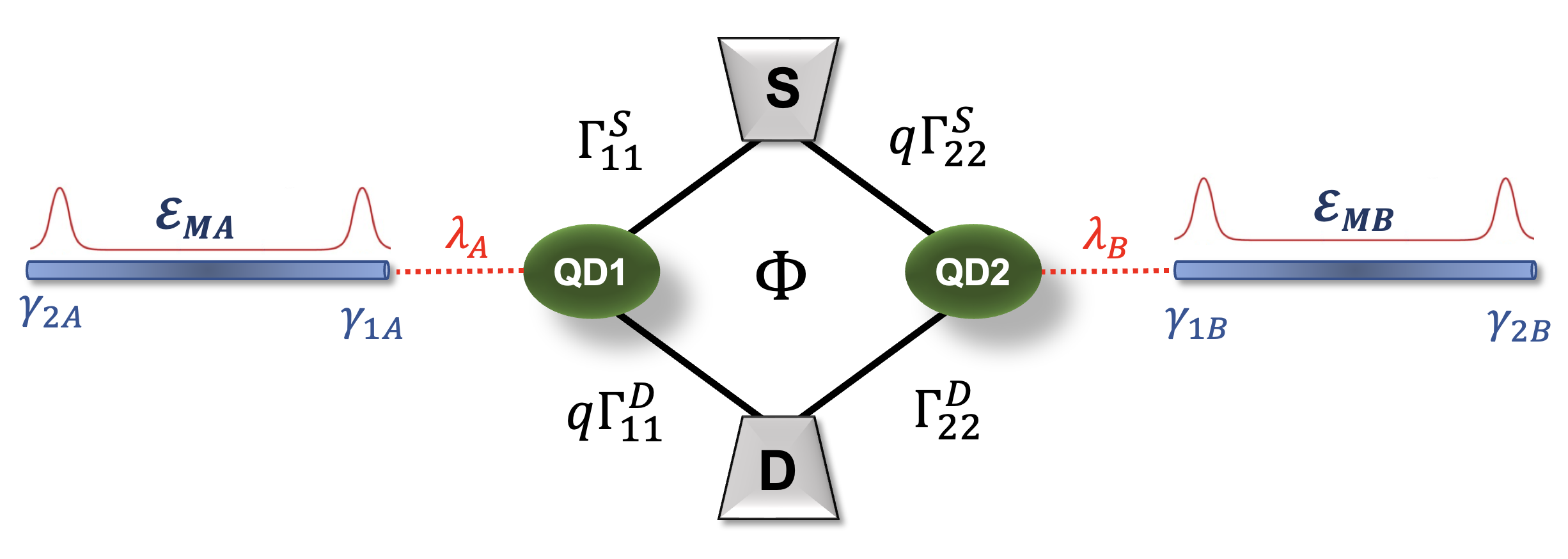}
    \caption{Schematic view of the system under study: TSCN-DQD-TSCN. Each QD (green) is coupled to a TSCN (blue tones). The TSCN $A(B)$ is connected to the QD1(2) and hosts two MZMs, $\gamma_{_{1,A(B)}}$ and $\gamma_{_{2,A(B)}}$, at its ends (light blue). The DQD is coupled to two normal leads, labeled as S and D (solid gray), and an external magnetic flux $\Phi$ across the interferometer is considered. The $\Gamma$ parameters are the couplings among the system's components and $q$ is an asymmetry parameter, as later explained in the main text.}
    \label{fig1}
\end{figure}

We consider an interferometer configuration of the DQD, where each QD is connected to the two normal leads S and D, and independently side-coupled to one of the TSCNs hosting MZMs at both ends, as we show schematically in the Fig.\ \ref{fig1}. We model the system through an effective low-energy Hamiltonian in the following form
%%%%%%%%%%%%%%%%%%%%%%%%%%%%%%% FIG.
%%%%%%%%%%%%%%%%%%%%%%%%%%%%%%%
\begin{equation}
    H = H_{\text{dots}}+H_{\text{leads}}+H_{\text{dots-leads}}+H_{\text{M}}+H_{\text{dots-M}}\,,\label{eq1}
\end{equation}
where the first three terms on the right-hand side correspond to the regular electronic contributions, given by
\begin{align}
    H_{\text{dot}}        &= \sum_{j=1,2}\varepsilon_{j}d^{\dagger}_{j}d_{j}\,,\label{2}\\
    H_{\text{leads}}      &= \sum_{\alpha,{\bf k}}\varepsilon_{\alpha,{\bf k}}a^{\dagger}_{\alpha,{\bf k}}a_{\alpha,{\bf k}}\,,\label{3}\\
    H_{\text{dots-leads}} &= \sum_{\alpha,{\bf k}}\sum_{j=1,2}(t_{\alpha,{\bf k},j}(\varphi_{\alpha, j})a^{\dagger}_{\alpha,{\bf k}}d_{j}+\text{h.c.})\,,\label{4}
\end{align}
where $d^{\dagger}_{j}(d_{j})$ is the electron's creation (annihilation) operator in the $j$-th QD, with single energy level $\varepsilon_{j}$. The operator $a^{\dagger}_{\alpha,{\bf k}}(a_{\alpha,{\bf k}})$ is the electron creation (annihilation) operator with momentum \textbf{k}, and energy $\varepsilon_{\alpha,{\bf k}}$ in the lead $\alpha=S,D$. The parameter $t_{\alpha, {\bf k},j}(\varphi_{ \alpha, j})=t_{\alpha,{\bf k},j}^{(0)}\exp{[i\varphi_{\alpha,j}]}$ describes the QD-lead tunnel matrix element, where an Aharonov-Bohm (AB) phase is included to model the magnetic flux $\Phi$ across the interferometer \cite{kubala2002flux}. We choose a symmetric gauge such that $\varphi_{_{D,1}}=-\varphi_{_{D,2}}=-\varphi_{_{S,1}}=\varphi_{_{S,2}}=\phi/4$, with $\phi=2\pi\Phi/\Phi_{0}$ and $\Phi_{0}=h/e$ is the quantum flux, where $h$ is the Planck's constant and $e$ the electron's charge.

The last two terms in the Hamiltonian presented in Eq.\ (\ref{eq1}) are MZMs terms, specifically MZM-MZM and MZM-QD couplings, given by
\begin{equation}
    H_{\text{M}} = \sum_{\beta}i\varepsilon_{_{M,\beta}}\gamma_{_{1,\beta}}\gamma_{_{2,\beta}}\label{5}\,,
\end{equation}
\begin{equation}
    H_{\text{dot-M}} = (\lambda_{A}d_{1}-\lambda^{*}_{A}d^{\dagger}_{1})\gamma_{_{1,A}}+(\lambda_{B}d_{2}-\lambda^{*}_{B}d^{\dagger}_{2})\gamma_{_{1,B}}\label{6}\,,
\end{equation}
where $\gamma_{_{j,\beta}}$ denotes the MZM operator (with $\beta=A,B$), and satisfies both $\gamma_{_{j,\beta}}=\left[\gamma_{_{j,\beta}}\right]^{\dagger}$, and $\{\gamma_{_{j,\beta}},\gamma_{_{j',\beta'}}\}=\delta_{j,j'}\delta_{\beta,\beta'}$. Besides, $\lambda_{A(B)}$ is the tunneling coupling between $\gamma_{1,A(B)}$ and the QD$_{1(2)}$, and  $\varepsilon_{M,\beta}\propto \exp{(-L_{\beta}/\zeta)}$ is the coupling amplitude between two MZMs in the same TSCN, where $L_{\beta}$ corresponds to the wire's length and $\zeta$ denotes the superconducting coherence length. We can evaluate the electronic transport by using a transformation as follows: by writing each MZM operator as a superposition of regular fermionic operators $f_{\beta}$ in the form
\begin{subequations}
    \begin{equation}
        \gamma_{_{1,\beta}} = \frac{1}{\sqrt{2}}(f_{\beta}+f^{\dagger}_{\beta})\label{7}\,,
    \end{equation}
    \begin{equation}
        \gamma_{_{2,\beta}} = -\dfrac{i}{\sqrt{2}}(f_{\beta}-f^{\dagger}_{\beta})\label{8}\,,
    \end{equation}
\end{subequations}
satisfying the anticommutation relations $\{f_{\beta},f_{\beta'}\}=\{f_{\beta}^{\dagger},f_{\beta'}^{\dagger}\}=0$, and $\{{f_{\beta},f_{\beta'}^{\dagger}}\}=\delta_{\beta,\beta'}$. Accordingly, the Eq.\ (\ref{5}) and Eq.\ (\ref{6}) transform to
\begin{equation}
    H_{\text{M}} = \sum_{S}\varepsilon_{_{M,\beta}}\left(f_\beta^{\dagger}f_{\beta}-\dfrac{1}{2}\right)\label{9}\,,
\end{equation}
\begin{eqnarray}
    H_{\text{dot-M}} &=& \left(\dfrac{1}{\sqrt{2}}\right)(\lambda_{A}d_{1}-\lambda^{*}_{A}d^{\dagger}_{1})(f_{A}+f^{\dagger}_{A})\nonumber \\
                     &+& \left(\dfrac{1}{\sqrt{2}}\right)(\lambda_{B}d_{2}-\lambda^{*}_{B}d^{\dagger}_{2})(f_{B}+f^{\dagger}_{B})\label{10}\,.
\end{eqnarray}

The Hamiltonian described above is spinless since only electrons with one spin projection will couple to the MZMs  \cite{ruiz2015interaction}. At low temperatures, characteristic of superconducting systems, the linear electronic conductance $\mathcal{G}$ is obtained through the transmission probability $T(\omega)$. We fixed the temperature at $\mathcal{T}=0$, so the relation between both quantities is directly given by the Landauer formula $\mathcal{G}=(e^2/h)T(\omega=\varepsilon_{\text{F}})$ \cite{datta2005quantum, meir1992landauer}, where $\varepsilon_{\text{F}}$ is the Fermi level's energy. The transmission probability is calculated from the expression
\begin{equation}
    T(\omega) = \text{Tr}\{\hat{G}^{a}(\omega)\hat{\Gamma}^{D}\hat{G}^{r}(\omega)\hat{\Gamma}^{S}\}\label{11} \,,  
\end{equation}
where $\hat{G}^{a(r)}(\omega)$ is the system advanced (retarded) GF in the energy domain, and $\hat{\Gamma}^{D(S)}$ the line-width function denoting the coupling between the QDs and the leads $D(S)$, and are given by
\begin{equation}
    \hat{\Gamma}^{\alpha}= \left(
    \begin{matrix}
        0 & 0 & 0                     & 0                     & 0                     & 0                     & 0 & 0 \\
        0 & 0 & 0                     & 0                     & 0                     & 0                     & 0 & 0 \\
        0 & 0 & \Gamma^{\alpha}_{11}  & 0                     & \Lambda^{\alpha}_{12} & 0                     & 0 & 0 \\
        0 & 0 & 0                     & \Gamma^{\alpha}_{11}  & 0                     & \Lambda^{\alpha}_{12} & 0 & 0 \\
        0 & 0 & \Lambda^{\alpha}_{21} & 0                     & \Gamma^{\alpha}_{22}  & 0                     & 0 & 0 \\
        0 & 0 & 0                     & \Lambda^{\alpha}_{21} & 0                     & \Gamma^{\alpha}_{22}  & 0 & 0 \\
        0 & 0 & 0                     & 0                     & 0                     & 0                     & 0 & 0 \\
        0 & 0 & 0                     & 0                     & 0                     & 0                     & 0 & 0
    \end{matrix}
    \right)
    \label{12}\,,
\end{equation}
where we have defined $\Lambda^{\alpha}_{ij}=\sqrt{\Gamma^{\alpha}_{ij}\Gamma^{\alpha}_{ji}}$, and $\Gamma^{\alpha}_{ij}=2\pi t_{\alpha,{\bf k},i}(\varphi_{ \alpha i})[t_{\alpha,{\bf k},j}(\varphi_{ \alpha j})]^{*}\rho_{\alpha}$ is the tunnel-coupling strength, with $\rho_{\alpha}$ being the local density of states in the lead $\alpha$. The retarded GF satisfies $\hat{G}^{r}(\omega)=[\hat{G}^{a}(\omega)]^{\dagger}$, and will be obtained by means of direct inversion, i. e.  $\hat{G}^{r}(\omega)=(\hat{\omega}-\hat{H})^{-1}$, where $\hat{\omega}=\omega \hat{I}$ is the energy matrix. The procedure is presented qualitatively in the Appendix\ \ref{AppendixA}.

We also investigate the behavior of the local density of states (LDOS) in each QD, since it is closely related to resonances in the conductance. The LDOS is expressed as
\begin{equation}
    \text{LDOS}_{1(2)}(\omega)=-\dfrac{1}{\pi}\text{Im}\left[G^{r}_{33(55)}(\omega)+G^{r}_{44(66)}(\omega)\right]\label{13}\,,
\end{equation}
where $G^{r}_{33(44)}\left(G^{r}_{55(66)}\right)$ are the diagonal terms of the full GF $\hat{G}^{r}(\omega)$ corresponding to the QD1(QD2). Finally, the spectral functions for the MZMs are given by 
\begin{equation}
    A_{A}(\omega)=-2\,\text{Im}\left[\sum_{i,j=1,2}G^{r}_{ij}(\omega)\right]\label{14A}\,,
\end{equation}
\begin{equation}
    A_{B}(\omega)=-2\,\text{Im}\left[\sum_{i,j=7,8}G^{r}_{ij}(\omega)\right]\label{14B}\,,
\end{equation}
where $G^{r}_{ij} (\omega)$ in Eq.\ (\ref{14A}) and Eq.\ (\ref{14B}) are matrix elements extracted from the full GF $\hat{G}^{r}(\omega)$, corresponding to the MZMs operators $\gamma_{_{1,A}}$ and $\gamma_{_{1,B}}$, respectively.

Moreover, the complete GF poles are closely related to the eigenvalues of the isolated TSCN-DQD-TSCN (disconnected from leads S and D), and give reliable information about the energy localization of the system's states. The eigenvalues can be written as
\begin{eqnarray}
    2[\omega_{1(2)}^{\pm}]^{2}&=&\varepsilon_{_{1(2)}}^{2}+\varepsilon_{_{M,A(B)}}^{2}+2\lambda_{A(B)}^{2} \nonumber \\
    &\pm& \sqrt{(\varepsilon_{_{1(2)}}+\varepsilon_{_{M,A(B)}})^{2}+2\lambda_{A(B)}^{2}}\nonumber \\
    &\times& \sqrt{(\varepsilon_{_{1(2)}}-\varepsilon_{_{M,A(B)}})^{2}+2\lambda_{A(B)}^{2}}\label{15ev}\,.
\end{eqnarray}

For the particular case of $\lambda_{A(B)}=\lambda$, $\varepsilon_{_{MA(B)}}=\varepsilon_{_{M}}$, and $\varepsilon_{_{1(2)}}=\varepsilon=0$, the eigenvalues $\omega_{1(2)}^{\pm}=\omega^{\pm}$ are
\begin{eqnarray}
    \omega^{-}=0\label{15omega}\,,
\end{eqnarray}
\begin{eqnarray}
    \omega^{+}=\pm\sqrt{\varepsilon_{_{M}}^{2}+2\lambda^{2}}\label{16omega} \,,
\end{eqnarray}
where $\omega^{-}$ in the Eq.\ (\ref{15omega}) has quadruple degeneracy and each solution for $\omega^{+}$ in the Eq.\ (\ref{16omega}) has double degeneracy.

%%%%%%%%%%%%%%%%%%%%%%%%%%%%%%%%%%%%%%%%%%%%%%%%%%%%%%%%%%%%%%
\section{Results}\label{secIII}
%%%%%%%%%%%%%%%%%%%%%%%%%%%%%%%%%%%%%%%%%%%%%%%%%%%%%%%%%%%%%%

We have considered the wide-band approximation, in which $\rho_{\alpha}$ has an approximately constant value and then $\Gamma^{\alpha}_{ii}$ is energy-independent. Thus, we fixed the line-width function to $\Gamma^{S}_{11}=\Gamma^{D}_{22}=\Gamma$, and $\Gamma^{D}_{12}=\Gamma^{S}_{12}=\Gamma^{D}_{21}=\Gamma^{S}_{21}=\Gamma^{D}_{11}=\Gamma^{S}_{22}=q\Gamma$, where $q$ is a dimensionless parameter with $q=1(q=0)$ corresponding to a close(open) system. Besides, the elements $\Gamma^{D}_{12}$, $\Gamma^{S}_{12}$, $\Gamma^{D}_{21}$, and $\Gamma^{S}_{21}$ contain the information of the AB phase due to magnetic flux. In the following, all the energy parameters are given in units of $\Gamma$. In order to consider realistic parameters with experiments, the values for $\Gamma$ can be considered from a few to hundreds of meV.

%%%%%%%%%%%%%%%%%%%%%%%%%%%%%%%
\subsubsection{Without magnetic flux ($\phi = 0$)}
%%%%%%%%%%%%%%%%%%%%%%%%%%%%%%%

%%%%%%%%%%%%%%%%%%%%%%%%%%%%%%% FIG.
\begin{figure}
    \centering
    \includegraphics{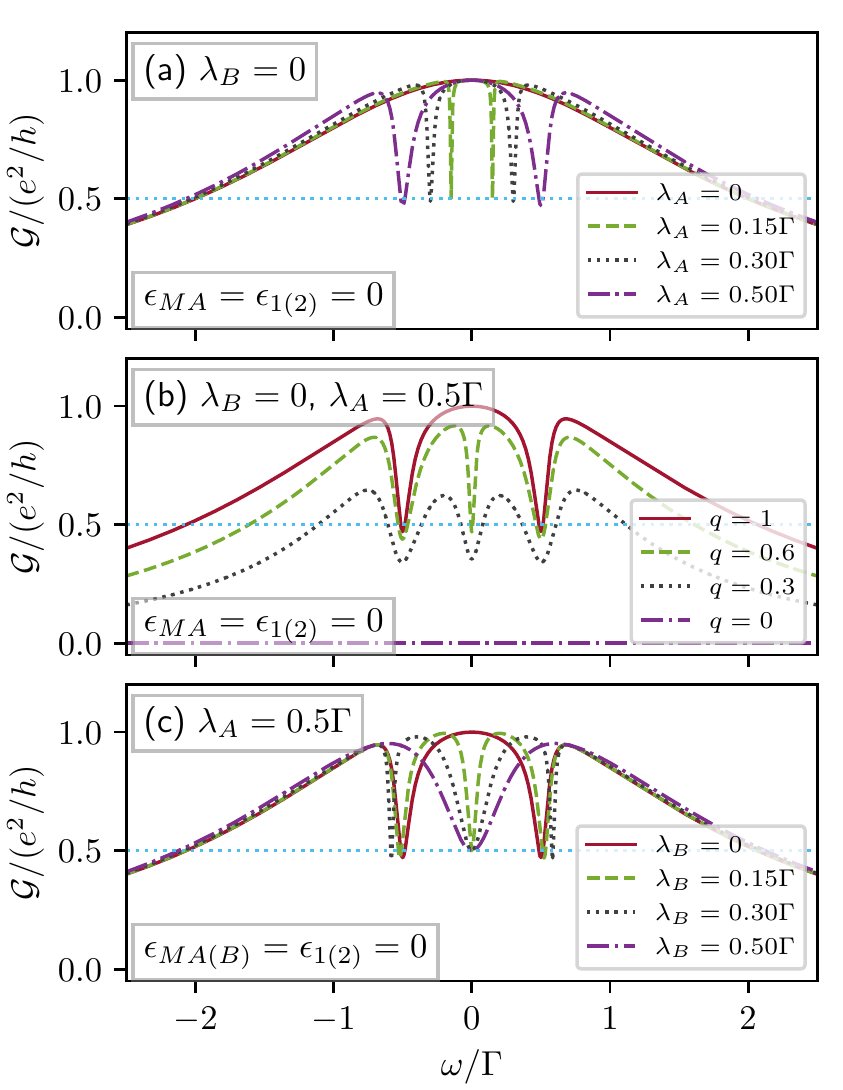}
    \caption{Linear conductance $\mathcal{G}$ as a function of the energy $\omega$ across the DQD in the long-wire limit for both TSCNs ($\varepsilon_{_{MA(B)}}=0$) with $\varepsilon_{_{1(2)}}=0$. Panel (a) shows $\mathcal{G}$ with one TSCN coupled ($\lambda_{A}\neq0$ and $\lambda_{B}=0$) where the solid red, dashed green, dotted black and dash-dotted magenta lines correspond to $\lambda_{A}/\Gamma=\{0, 0.15, 0.3, 0.5\}$, respectively. Panel (b) shows $\mathcal{G}$ using fixed $\lambda_{A}=0.5\Gamma$ and $\lambda_{B}=0$ for the asymmetric coupling between contacts-QD $(0 \leq q \leq 1)$. Panel (c) shows $\mathcal{G}$ when both TSCNs are connected ($\lambda_{A}=0.5\Gamma$ and $\lambda_{B}\neq0$) using symmetric coupling  between contacts-QD $(q=1)$, where the solid red, dashed green, dotted black and dash-dotted magenta lines correspond to the values $\lambda_{B}/\Gamma=\{0, 0.15, 0.3, 0.5\}$, respectively.}
    \label{fig2}
\end{figure}
%%%%%%%%%%%%%%%%%%%%%%%%%%%%%%%

%%%%%%%%%%%%%%%%%%%%%%%%%%%%%%% FIG.
\begin{figure}[h]
    \centering
    \includegraphics{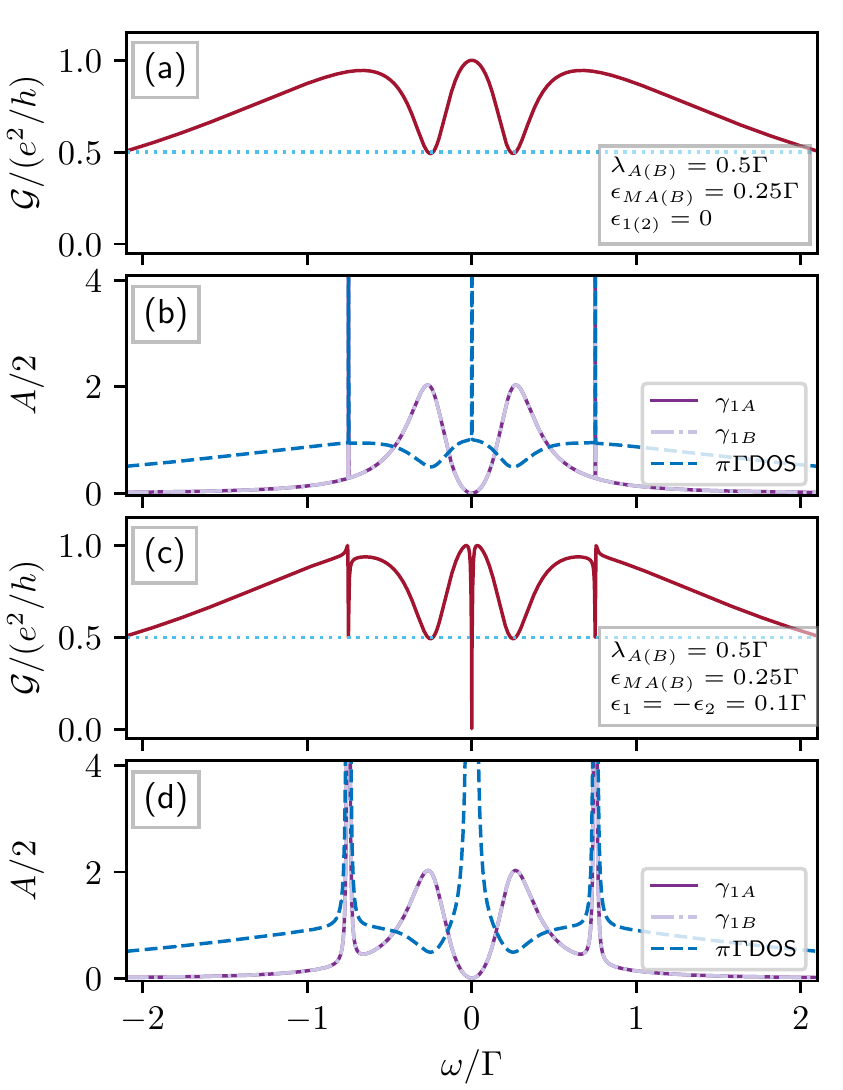}
    \caption{Linear conductance $\mathcal{G}$, spectral function $A/2$ and DOS as a function of the energy $\omega$ across the DQD out of the long-wire limit for both TSCNs ($\lambda_{A(B)}=0.5\Gamma$ and $\varepsilon_{MA(B)}=0.25\Gamma$). Panels (a) and (c) show $\mathcal{G}$ with $\varepsilon_{_{1(2)}}=0$ and $\varepsilon_{_{1}}=-\varepsilon_{_{2}}=0.1\Gamma$, respectively. Panels (b) and (d) show the QDs' spectral function $A/2$ and DOS for the cases described in panels (a) and (c), respectively.}
    \label{fig3}
\end{figure}
%%%%%%%%%%%%%%%%%%%%%%%%%%%%%%%

First, we consider the case with both TSCNs in the long-wire limit using $\varepsilon_{_{MA(B)}}=0$, and we fix the QDs' energy levels at $\varepsilon_{_{1(2)}}=0$. Fig.\ \ref{fig2} shows the linear conductance $\mathcal{G}$, as a function of the energy $\omega$, for the case when one or both TSCNs are coupled to the DQD with $q=1$ $(q=0)$ corresponding to a close (open) system. Figure \ref{fig2}(a) shows a Breit-Wigner resonance centered at $\omega=0$ for $\lambda_{A}=0$ (solid red line) while for $\lambda_{A}\neq0$ the linear conductance is composed of a maximum at $\omega=0$ and two dips located at energies $\omega=\pm \lambda_{A}$, the latter reaching conductance values $\mathcal{G} \approx e^{2}/2h$. In Fig. \ref{fig2}(b), a fixed $\lambda_{A}=0.5\Gamma$ is used, and the two QDs are connected in a parallel configuration with asymmetrical left-right coupling such as $0 \leq q \leq 1$. The linear conductance $\mathcal{G}$ falls progressively to zero when the circuit goes from the close ($q=1$, solid red line) to the open system ($q=0$, dashed-dotted magenta line). In Fig.\ \ref{fig2}(c) we set the coupling strength $\lambda_{A}=0.5\Gamma$ and the second TSCN is connected allowing $0 \leq \lambda_{B} \leq 0.5\Gamma$. The conductance exhibits two antiresonances close to the conductance value $\mathcal{G}\approx e^2/2h$, which are located at energies $\omega \approx \pm \tilde{\lambda}\sqrt{2}$, with $\tilde{\lambda}=\lambda_{A}+\lambda_{B}$ whenever $\lambda_{A}\neq\lambda_{B}$. These vanish for a symmetrical coupling strength $\lambda_{A}=\lambda_{B}=0.5\Gamma$ due to the MZMs hybridization. Furthermore, at zero energy ($\omega=0$) the conductance goes from its maximum value $\mathcal{G}=e^2/h$ (with one TSCN coupled) to a half-maximum value $\mathcal{G}=e^2/2h$  (when both TSCNs are coupled to the DQD). This robust behavior is independent of the coupling strength $\lambda_{A}$ and $\lambda_{B}$, the latter being a MZM signature as was reported for the first time by Liu and Baranger \cite{liu2011detecting}. Without loss of generality, we now set $q=1$ in what follows.

%%%%%%%%%%%%%%%%%%%%%%%%%%%%%%% FIG.
\begin{figure*}
    \centering
    \includegraphics[width=26em,height=2.9em]{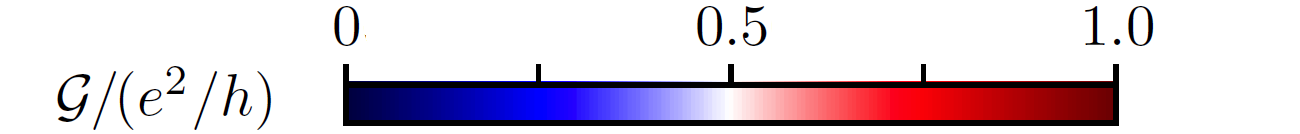}
    \includegraphics[width=54em,height=20em]{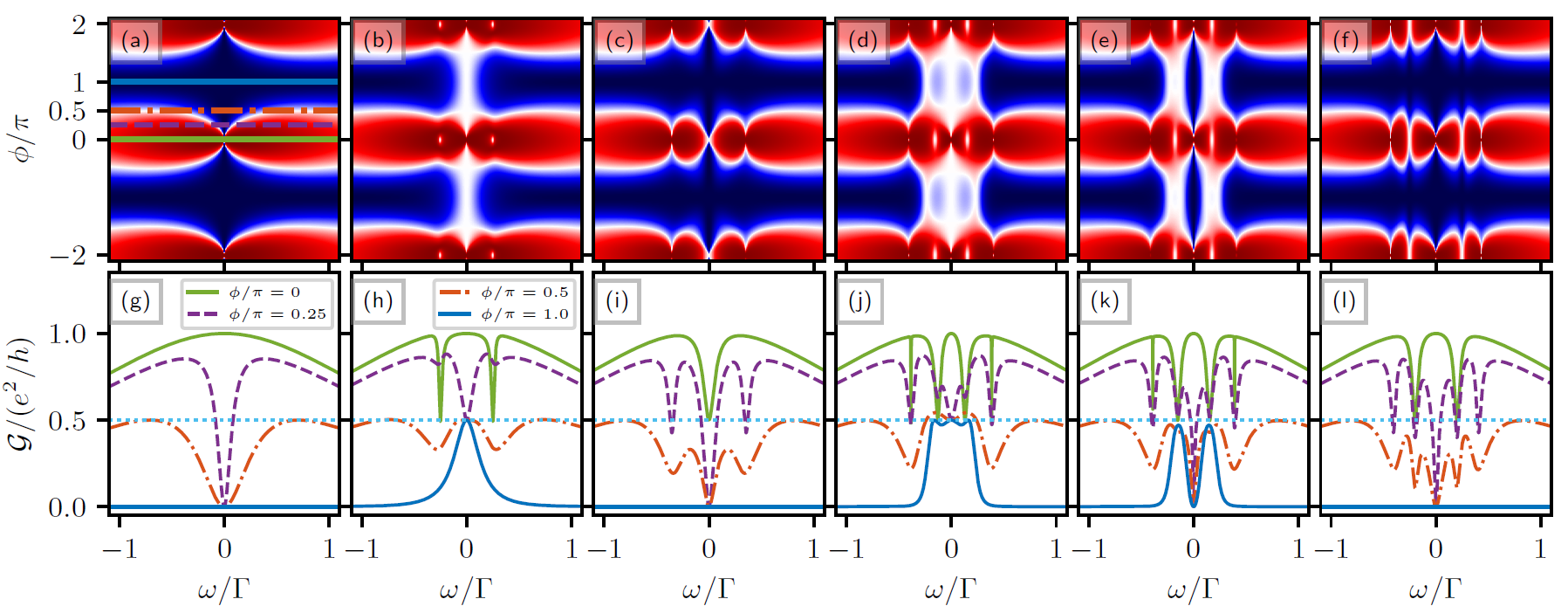}
    \caption{Color map of the linear conductance $\mathcal{G}$ as a function of both the magnetic flux $\phi$ and the energy $\omega$ in upper panels (a)-(f), where the red(blue) color represents a maximum(minimum) value. $\mathcal{G}$ as a function of the energy $\omega$ for different values of the magnetic flux in lower panels (g)-(l), where the solid green, dashed magenta, dash-dotted orange, and solid blue lines correspond to $\phi/\pi=\{0, 0.25, 0.5, 1\}$, respectively. The QDs' energy levels are $\varepsilon_{_{1(2)}}=0$ in all panels. In panels (a) and (g) $\lambda_{A(B)}=0$ is used. In panels (b) and (h) $\lambda_{A}=0.25\Gamma$, $\varepsilon_{_{MA}}=0$, and $\lambda_{B}=0$ are used. In panels (c) and (i) $\lambda_{A(B)}=0.25\Gamma$, and $\varepsilon_{_{MA(B)}}=0$ are used. In panels (d) and (j) $\lambda_{A(B)}=0.25\Gamma$, $\varepsilon_{_{MA}}=0.2\Gamma$, and $\varepsilon_{_{MB}}=0$ are used. In panels (e) and (k) $\lambda_{A(B)}=0.25\Gamma$, $\varepsilon_{_{MA}}=0.2\Gamma$, and $\varepsilon_{_{MB}}=0.1\Gamma$ are used. In panels (f) and (l) $\lambda_{A(B)}=0.25\Gamma$, and $\varepsilon_{_{MA(B)}}=0.2\Gamma$ are used.}
    \label{fig:wide4}
\end{figure*}
%%%%%%%%%%%%%%%%%%%%%%%%%%%%%%%

Figure\ \ref{fig3} considers both TSCNs out of the long-wire limit  $(i. e.$ $  \varepsilon_{_{MA(B)}}=\varepsilon_{_{M}}=0.25\Gamma)$ and the coupling strength of each TSCN-QD connection as $\lambda_{A(B)}=\lambda=0.5\Gamma$. The linear conductance $\mathcal{G}$ as a function of the energy $\omega$, is shown in Fig.\ \ref{fig3}(a)-\ref{fig3}(c), and the spectral function $A(\omega)/2$ with the DQD's DOS is shown in Fig.\ \ref{fig3}(b)-\ref{fig3}(d). The DOS is obtained by adding both contributions given by Eq.\ (\ref{13}). In the case when the QDs' energy levels $\varepsilon_{_{1(2)}}=0$, the linear conductance $\mathcal{G}$ is composed of a maximum in $\omega=0$ and two dips located at energies $\omega=\pm \varepsilon_{_{M}}$ [Fig.\ \ref{fig3}(a)]. In Fig.\ \ref{fig3}(b), localized states are observed in the DOS (dashed blue line) in the form of resonances with vanishing width. Since they do not have a projection in $\mathcal{G}$, these states correspond to BICs placed at energies $\omega=\omega^{-}=0$ (quadruply degenerate) and  $\omega=\omega^{+}=\pm\sqrt{\varepsilon_{_{M}}^{2}+2\lambda^{2}}$ (each doubly degenerate), accordingly to the eigenvalues described in Eq.\ (\ref{15omega}) and Eq.\ (\ref{16omega}), respectively. Note that the side BICs observed in the DOS are related to the leaking of the lateral bound states present in the spectral function $A(\omega)$. The two symmetric broad resonances placed at energies $\omega=\pm \varepsilon_{_{M}}$ in $A(\omega)$ hybridize with DQD states, giving place to the dips in the linear conductance. In this case, the spectral function of each MZM, $A_{A}(\omega)$ and $A_{B}(\omega)$, are the same and are described by the solid magenta line and the dashed-dotted light-magenta line.

In Fig.\ \ref{fig3}(c), we introduce asymmetry in the QDs' energy levels in the form $\varepsilon_{_{1}}=-\varepsilon_{_{2}}=0.1\Gamma$. The linear conductance $\mathcal{G}$ shows an antiresonance located at $\omega=\omega^{-}=0$, and shows two asymmetric Fano-like antiresonances located at the eigenvalues $\omega=\omega^{+}$, given by the Eq.\ (\ref{15ev}). These states observed in the spectral function and the DOS in Fig.\ \ref{fig3}(d) acquire a width, becoming quasi-BICs since they acquire projections on the linear conductance. The vanishing of the lateral Fano-like shapes in both the symmetrical energy and the TSCNs-coupled cases, termed as the so-called Ghost-Fano Majorana effect, is a direct consequence of the bound states uncoupling from the rest of the system, and then there is no contribution to the transmission coefficient \cite{de2003ghost}.

%%%%%%%%%%%%%%%%%%%%%%%%%%%%%%%
\subsubsection{With magnetic flux ($\phi \neq 0$)}
%%%%%%%%%%%%%%%%%%%%%%%%%%%%%%%

\begin{figure}[h!]
    \centering   \includegraphics{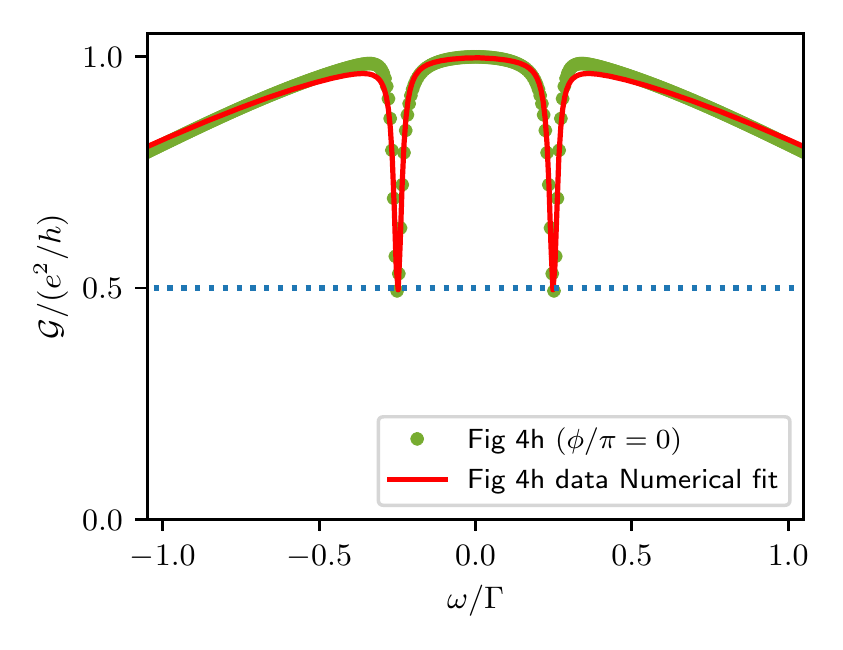}
    \caption{Conductance $\mathcal{G}$ from Fig. \ref{fig:wide4}(h) for $\lambda_{A}=0.25\Gamma$, $\lambda_{B}=0$, $\varepsilon_{_{MA}}=0$, and $\phi=0$. The solid red line is a numerical fit using Eq. (\ref{eqfitteo}), with $\delta = 0.249$, $\xi = 0.014$, and $\eta = 2.142$ (all in energy units of $\Gamma$).}
    \label{figfitteo}
\end{figure}

We study the electronic transport in the TSCN-DQD-TSCN system in the presence of a magnetic flux across the interferometer $(\text{so }\phi\neq0)$. Figure\ \ref{fig:wide4} shows the color map of the linear conductance $\mathcal{G}$ as a function of the dimensionless magnetic flux $\phi$ and the energy $\omega$ in top panels. Each bottom panel is a linear conductance $\mathcal{G}$'s horizontal cut (of the corresponding top panel) at fixed values of the magnetic flux $\phi/\pi=\{0, 0.25, 0.5, 1\}$, presented in green, magenta, orange, and blue colors, respectively. We use fixed QDs' energy levels $\varepsilon_{_{1(2)}}=0$ in all panels. Figure\ \ref{fig:wide4}(a) corresponds to a DQD interferometer (without TSCNs, $\lambda_{A(B)}=0)$ where the magnetic flux induces a transport suppression for a wide range of values ($\mathcal{G}=0$, blue color in the color maps). Kubala and K\"onig described this behavior  \cite{kubala2002flux}, where the QDs' energy levels are coupled to each other indirectly via the leads. This coupling yields signatures of a flux-dependent level attraction in the linear conductance. Besides, transport suppression occurs whenever both QD levels are close to the Fermi level of the leads. Coupling one TSCN in the long-wire limit $(\lambda_{A}=0.25\Gamma$; $\varepsilon_{_{MA}}=0)$ allows transmission around the energy $\omega=0$, and is independent of the magnetic flux. A half-integer conductance describes this behavior in Fig.\ \ref{fig:wide4}(b), and is also shown in Fig.\ \ref{fig:wide4}(h), where the linear conductance at zero energy takes $\mathcal{G}=e^{2}/2h$ values when $\phi \neq 2\pi n$ (with integer $n$). For the case $\phi=0$, shown in Fig.\ \ref{fig:wide4}(h), we can characterize the linear conductance by means of a convolution of a Fano and a Breit-Wigner line shapes in the form
\begin{equation}
    F (\omega) = \frac{|\varepsilon+\tilde{q}|^2}{\varepsilon^2+1} \frac{\eta^2}{\omega^2+\eta^2}\,,\label{eqfitteo}
\end{equation}
where we have used $\varepsilon=(|\omega|-\delta)/\xi$, and the complex $\tilde{q}$-parameter $\tilde{q}=i/\sqrt{2}$. The latter can be interpreted as an evidence of the presence of a  superconductor lead \cite{calle2020fano}, and the particular pure imaginary value seems to be the signature of the leaking of the MZM into the QDs. The $\delta$ parameter describes the localization of the antiresonance and, in this particular case, can be identified as the coupling strength $\lambda_A$. The comparison between the exact result and the fitting using Eq.\ (\ref{eqfitteo}) is shown in Fig.\ \ref{figfitteo}. 

By symmetrically coupling both TSCNs in the long-wire limit $(\lambda_{A(B)}=0.25\Gamma$; $\varepsilon_{_{MA(B)}}=0)$ we obtain a transport suppression as a function of the magnetic flux as shown in Fig.\ \ref{fig:wide4}(c), same as previously in Fig.\ \ref{fig:wide4}(a). Additionally, when $\omega=0$ the linear conductance reaches a half-integer value ($\mathcal{G}=e^{2}/2h$) for magnetic flux values $\phi=2n\pi$, as is also shown in Fig.\ \ref{fig:wide4}(i), where the linear conductance takes the value $\mathcal{G}=e^{2}/2h$ for $\phi=0$ (green line), and takes the value $\mathcal{G}=0$ for $\phi \neq 0$ (magenta, orange and blue lines). 

%%%%%%%%%%%%%%%%%%%%%%%%%%%%%%% FIG.
\begin{figure}[h!]
    \centering
    \includegraphics{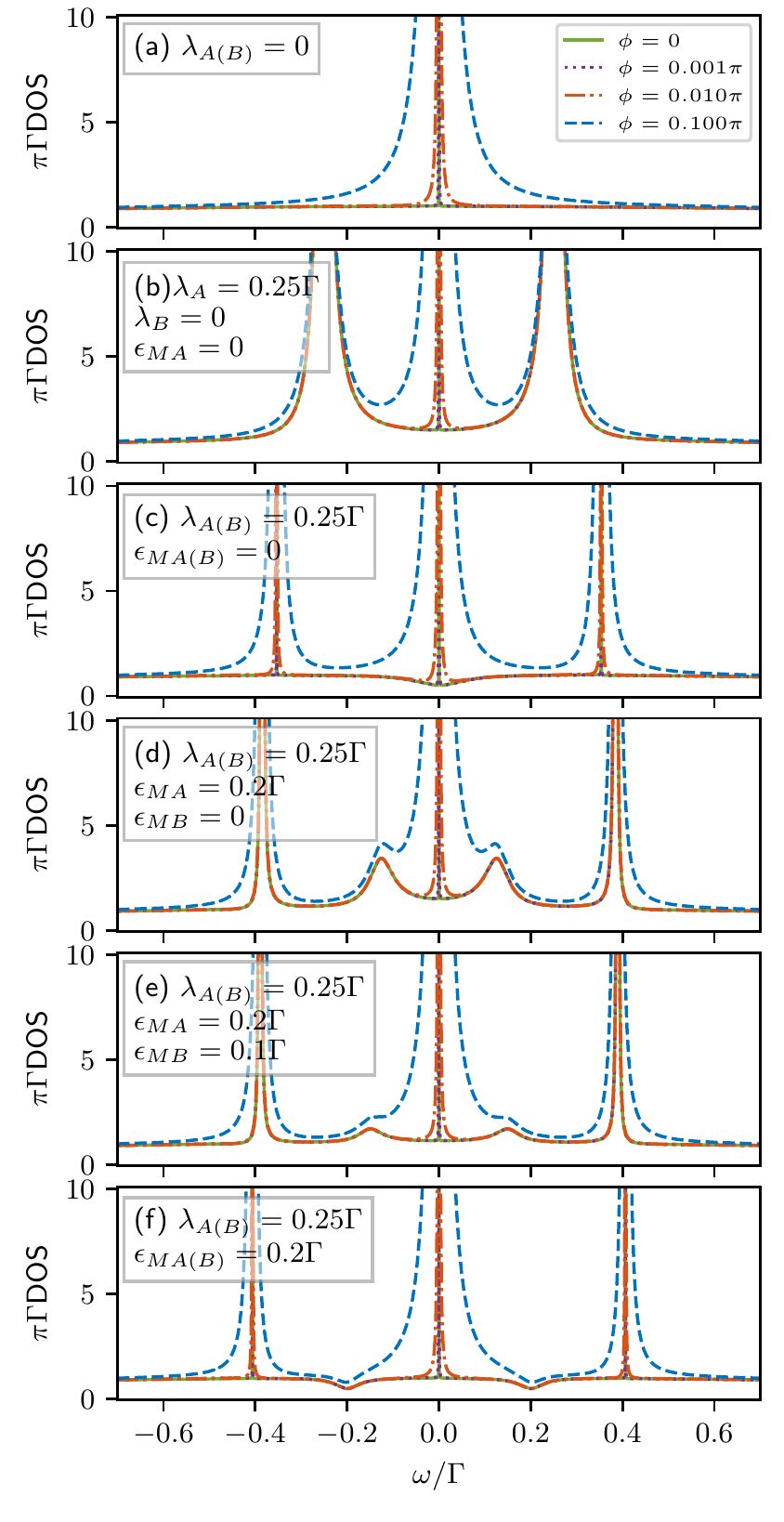}
    \caption{Density of states (DOS) as a function of the energy $\omega$ for different values of the magnetic flux, where the green, magenta, orange, and blue lines correspond $\phi=\{0, \pi/1000, \pi/100, \pi/10\}$, respectively. The QDs' energy levels are $\varepsilon_{1(2)}=0$. In panel (a) $\lambda_{A(B)}=0$ is used. In panel (b) $\lambda_{A}=0.25\Gamma$, $\varepsilon_{_{MA}}=0$, and $\lambda_{B}=0$ are used. In panel (c) $\lambda_{A(B)}=0.25\Gamma$, and $\varepsilon_{_{MA(B)}}=0$ are used. In panel (d) $\lambda_{A(B)}=0.25\Gamma$, $\varepsilon_{_{MA}}=0.2\Gamma$, and $\varepsilon_{_{MB}}=0$ are used. In panel (e) $\lambda_{A(B)}=0.25\Gamma$, $\varepsilon_{MA}=0.2\Gamma$, and $\varepsilon_{MB}=0.1\Gamma$ are used. In panel (f)  $\lambda_{A(B)}=0.25\Gamma$, and $\varepsilon_{_{MA(B)}}=0.2\Gamma$ are used.}
    \label{fig5}
\end{figure}
%%%%%%%%%%%%%%%%%%%%%%%%%%%%%%%

In the same way, in the Fig.\ \ref{fig:wide4}(d) we coupled both TSCNs considering one of them away from the long-wire limit $(\lambda_{A(B)}=0.25\Gamma$; $\varepsilon_{_{MA}}=0.2\Gamma$ and $\varepsilon_{_{MB}}=0)$, where the linear conductance presented in the color map shows that the system allows transport around $\omega=0$ as a function of the magnetic flux. The linear conductance as a function of the energy $\omega$ in Fig.\ \ref{fig:wide4}(j) shows that, for $\omega=0$, the maxima reached are $\mathcal{G}=e^{2}/2h$ for magnetic flux $\phi\neq2n\pi$ and $\mathcal{G}=e^{2}/h$ for $\phi=2n\pi$. Coupling both TSCNs away from the long-wire limit using different wires lengths $(\lambda_{A(B)}=0.25\Gamma$; $\varepsilon_{_{MA}}=0.2\Gamma$ and $\varepsilon_{_{MB}}=0.1\Gamma)$ is addressed in Fig.\ \ref{fig:wide4}(e). For this case we obtain two half-integer peaks in the linear conductance ($\mathcal{G}=e^{2}/2h$) for $\phi = 2n\pi$. When coupling both TSCNs with equal length ($\lambda_{A(B)}=0.25\Gamma$ and $\varepsilon_{_{MA(B)}}=0.2\Gamma$) as shown in Fig. \ref{fig:wide4}(f), the transport suppression as a function of the energy $\omega$ is recovered as in Fig. \ref{fig:wide4}(a) and Fig. \ref{fig:wide4}(c). For vanishing magnetic flux $(\phi=0)$, we obtain an integer maximum linear conductance $\mathcal{G}=e^{2}/h$ at zero energy, as well as two dips, reaching the half-integer value $\mathcal{G}=e^{2}/2h$, placed at energies $\omega=\varepsilon_{_{MA(B)}}=\pm 0.2\Gamma$, as is also shown in Fig.\ \ref{fig:wide4}(l) in the solid green line. 

We find that the magnetic flux induces a symmetry breaking in the DQD channels. This effect is described through zones of total reflection in the color map of the linear conductance (blue values of $\mathcal{G}=0$). In addition, however, we find that the device allows electronic transport in those zones of total reflection whenever the DQD has one of the following coupling cases: by coupling only one of the TSCN is in the long wire limit [Fig.\ \ref{fig:wide4}(b)]; by connecting both TSCNs with one of them in the long-wire limit [Fig.\ \ref{fig:wide4}(d)]; or by coupling both TSCNs out of the long-wire limit using different lengths in each one [Fig.\ \ref{fig:wide4}(e)], where the signal reaches the $\mathcal{G}=e^{2}/2h$ value. We interpret the latter signal as MZMs leaking into the QDs. Therefore, tuning the magnetic flux allows control of this leaking and, consequently, the possibility of manipulating the emergence of the MZMs signatures.

%%%%%%%%%%%%%%%%%%%%%%%%%%%%%%% FIG.
\begin{figure}
    \centering
    \includegraphics{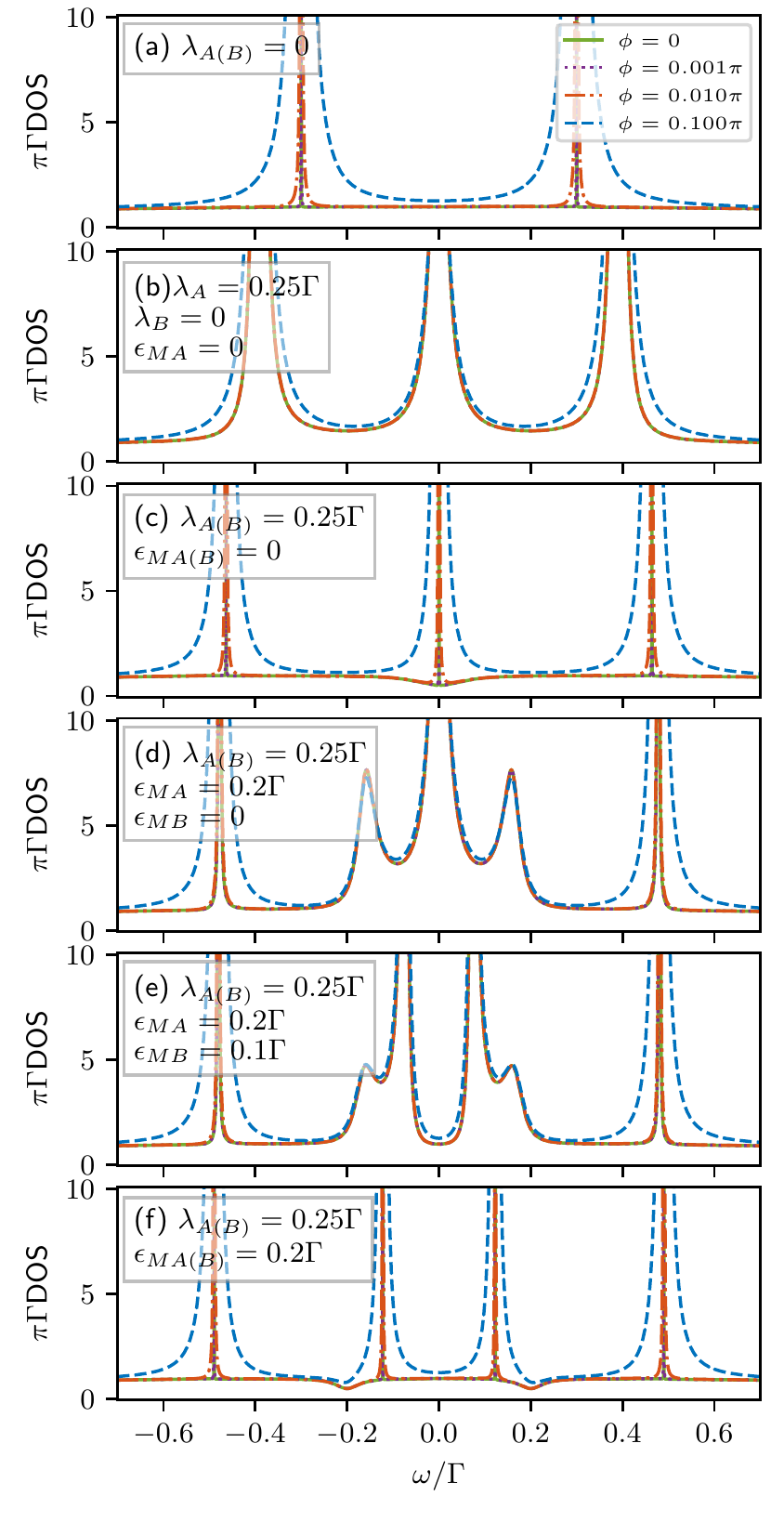}
    \caption{Density of states (DOS) as a function of the energy $\omega$ for different values of the magnetic flux, where the green, magenta, orange, and blue lines correspond to the following magnetic flux values: $\phi=\{0, \pi/1000, \pi/100, \pi/10\}$. These are the same calculations as the ones shown in Fig.\ \ref{fig5}, but now using QDs' energies $\varepsilon_{_{1(2)}}=0.3\Gamma$.}
    \label{fig6}
\end{figure}
%%%%%%%%%%%%%%%%%%%%%%%%%%%%%%%

Figure\ \ref{fig5} displays the density of states (DOS) as a function of the energy $\omega$ for magnetic fluxes $\phi$:  $0,$ $\pi/1000,$ $\pi/100,$ and $\pi/10$, in green, magenta, orange, and blue colors, respectively. The calculations in panels Fig.\ \ref{fig5}(a) to Fig.\ \ref{fig5}(f), use exactly the same combinations of parameters $\lambda_{A}$, $\lambda_{B}$, $\varepsilon_{_{MA}}$, and $\varepsilon_{_{MB}}$, as the bottom panels of Fig.\ \ref{fig:wide4} [Fig.\ \ref{fig:wide4}(g) to Fig.\ \ref{fig:wide4}(l)], correspondingly. In Fig.\ \ref{fig5}(a), we observe a resonance localized in $\omega=0$ for $\phi=0$, corresponding to a BIC, which width increases with the magnetic flux. We connected one TSCN in the long-wire limit in Fig.\ \ref{fig5}(b). We can also observe the BIC localized at $\omega=0$ for $\phi=0$. Besides, the DOS shows two symmetric wide resonances, centered at energies $\omega=\pm \lambda_{A}$, due to the hybridization of the MZMs with the QD, which are precisely the position of the antiresonance projected in the linear conductance, as shown in the green curve of Fig.\ \ref{fig:wide4}(h). Fig.\ \ref{fig5}(c) shows the DOS considering that the two TSCNs in the long-wire limit are connected. We observe the BIC placed at $\omega=0$ for $\phi=0$, and two symmetric BICs localized at $\omega^{+}=\pm \lambda_{A(B)}\sqrt{2}=0.25\sqrt{2}\Gamma$, which are given by taking $\varepsilon_{_{MA(B)}}=0$ in Eq. (\ref{16omega}). These states correspond to the hybridized MZMs, which do not show an apparent projection in the linear conductance [see Fig.\ \ref{fig:wide4}(i)]. We find that the central BIC in the DQD  at zero magnetic flux without  TSCNs is destroyed when only one TSCN is coupled to the system. However, it reappears when the two  TSCNs are connected. The above can be explained because the symmetry is restored at zero magnetic flux with two TSCNs coupled to the DQD. In a threaded magnetic flux, the central BIC becomes a quasi-BIC. When either one TSCN is out of the long-wire limit [as $\varepsilon_{_{MA}}=0.2\Gamma$ and $\varepsilon_{_{MB}}=0$ in Fig.\ \ref{fig5}(d)] or when both are out of the long-wire limit and also have different lengths [as $\varepsilon_{_{MA}}=0.2\Gamma$ and $\varepsilon_{_{MB}}=0.1\Gamma$ in Fig.\ \ref{fig5}(e)], we can observe that the two lateral symmetrical BICs in the DOS gain a finite width, hence becoming quasi-BICs. This can also be observed by projections in the linear conductance of the form of antiresonance values up to $\mathcal{G}=e^{2}/2h$ (Fig.\ \ref{fig:wide4}(j) and Fig.\ \ref{fig:wide4}(k), respectively) localized at the same energy. In the case of both TSCNs with finite and equal length [Fig.\ \ref{fig5}(f)], we recovered the two lateral symmetrical BICs placed at $\omega^{+}=\pm \sqrt{\varepsilon_{_{MA(B)}}^{2}+2\lambda_{A(B)}^{2}}=0.406\Gamma$ according with Eq. (\ref{16omega}). However, these states do not show projections in the linear conductance presented in Fig.\ \ref{fig:wide4}(l).

Figure\ \ref{fig6} shows the DOS for the same parameters of Fig.\ \ref{fig5}, but now breaking the symmetry in the QD' levels in the form of $\varepsilon_{_{1(2)}}=\varepsilon=0.3\Gamma$. In the case of the DQD without TSCNs [Fig.\ \ref{fig6}(a)], we can see two lateral peaks, corresponding to BICs located at energies $\omega=\pm \varepsilon_{_{1(2)}}=\pm 0.3\Gamma$. When one TSCN in the long-wire limit is connected $(\lambda_{A}=0.25\Gamma$, $\varepsilon_{_{MA}}=0)$, the DOS displays three wide resonances [Fig.\ \ref{fig6}(b)]. When connecting both TSCNs in the long-wire limit $(\lambda_{A(B)}=\lambda=0.25\Gamma$, $\varepsilon_{_{MA(B)}}=0)$ we obtain three resonances according to Eq.\ (\ref{15ev}), one located at $\omega^{-}=0$ (quadruply degenerate)  and two laterals located at $\omega^{+}=\pm\sqrt{\varepsilon^{2}+2\lambda^{2}}=\pm0.464\Gamma$ (each doubly degenerate). When one of the TSCN is taken out of the long-wire limit (Fig.\ \ref{fig6}(c) but adding $\varepsilon_{_{MA}}=0.2\Gamma)$, the BIC located at $\omega=0$ becomes a resonance. Also, the lateral resonances acquire a width due to the symmetry breaking [Fig.\ \ref{fig6}(d)], becoming quasi-BICs. In the same way, when both TSCNs are out of the long-wire limit $(\lambda_{A(B)}=0.25\Gamma$, $\varepsilon_{_{MA}}=0.2\Gamma$, and $\varepsilon_{_{MB}}=0.1\Gamma)$, the DOS exhibits two lateral quasi-BICs, and two broad lateral resonances located near to $\omega \approx 0$ [Fig.\ \ref{fig6}(e)]. In the symmetric case ($\lambda_{A(B)}=\lambda=0.25\Gamma$, $\varepsilon_{_{MA(B)}}=\varepsilon_{_{M}}=0.2\Gamma$), we can observe the formation of four BICs [Fig.\ \ref{fig6}(f)]. According to Eq.\ (\ref{15ev}), they are located at energies $\omega^{-}=\pm 0.122\Gamma$ and $\omega^{+}=\pm 0.490\Gamma$, and each resonance corresponds to a doubly degenerate state. We can interpret from the DOS that the formation of BICs occurs only in symmetric cases, as shown in Fig.\ \ref{fig6}(a), Fig.\ \ref{fig6}(c), and Fig.\ \ref{fig6}(f). Whenever one or both TSCNs are taken out of the long-wire limit (with different lengths), the lateral BICs disappear and become quasi-BICs [Fig.\ \ref{fig6}(d) and Fig.\ \ref{fig6}(e)].

%%%%%%%%%%%%%%%%%%%%%%%%%%%%%%%%%%%%%%%%%%%%%%%%%%%%%%%%%%%%%%
\section{Summary}\label{secIV}
%%%%%%%%%%%%%%%%%%%%%%%%%%%%%%%%%%%%%%%%%%%%%%%%%%%%%%%%%%%%%%

We studied a system formed by a DQD coupled to two normal leads forming an interferometer configuration. Each QD is independently side coupled to a TSCN hosting MZMs at both ends, and an external magnetic flux across the enclosed interferometer was considered. We focused on the linear conductance across the leads and the DOS of the system, obtained from both the MZMs spectral function and QDs' local density of states. The latter was obtained employing Green's function formalism. We show that the total reflection phenomenon is robust against the coupling of superconducting wires as long as the system's symmetry is maintained. Besides, for values of the magnetic flux $\phi = 2n\pi$, we find the formation of BICs due to the presence of the coupled MZMs, characterized through resonances of zero width in the DOS, in the same two favorable symmetric coupling cases mentioned above. However, as these states do not project in the conductance, behave as Ghost Fano Majorana anomalies. Also, we find that these BICs destroy as a function of the magnetic flux. On the other hand, whenever $\phi \neq 0$, these states acquire a finite width evolving to quasi-BICs and show a projection in the linear conductance in the form of antiresonances placed at the same energies. These results indicated we could control the bound states generated by switching this external parameter.

%%%%%%%%%%%%%%%%%%%%%%%%%%%%%%%%%%%%%%%%%%%%%%%%%%%%%%%%%%%%%%
\begin{acknowledgments}
A.P.G. is grateful for the funding of scholarship ANID-Chile No. 21210410. D.Z. acknowledges support from USM-Chile under Grant PI-LIR-2022-13, P.A.O. acknowledges support from FONDECYT grants 1201876 and 1220700.  
\end{acknowledgments}
%%%%%%%%%%%%%%%%%%%%%%%%%%%%%%%%%%%%%%%%%%%%%%%%%%%%%%%%%%%%%%

\section*{DATA AVAILABILITY STATEMENT}

Data will be made available on reasonable request.

%%%%%%%%%%%%%%%%%%%%%%%%%%%%%%%%%%%%%%%%%%%%%%%
\section*{CONFLICTS OF INTEREST}

The authors declare that they have no conflict of interest.

%%%%%%%%%%%%%%%%%%%%%%%%%%%%%%%%%%%%%%%%%%%%%%%
\section*{AUTHOR CONTRIBUTION STATEMENT}

All authors contributed equally and significantly in writing this article. All authors read and approved the final manuscript.

\onecolumngrid
%%%%%%%%%%%%%%%%%%%%%%%%%%%%%%%%%%%%%%%%%%%%%%%%%%%%%%%%%%%%%%
\appendix
%%%%%%%%%%%%%%%%%%%%%%%%%%%%%%%%%%%%%%%%%%%%%%%%%%%%%%%%%%%%%%
\section{Retarded Green function}\label{AppendixA}
%%%%%%%%%%%%%%%%%%%%%%%%%%%%%%%%%%%%%%%%%%%%%%%%%%%%%%%%%%%%%%

The retarded Green's function $\hat{G}^{r}(\omega)=[\hat{G}^{a}(\omega)]^{\dagger}$ is obtained by means of direct inversion, i.e. $\hat{G}^{r}(\omega)=(\hat{\omega}-\hat{H})^{-1}$ where $\hat{\omega}=\omega \hat{I}$ is the diagonal energy and $\hat{H}$ the system Hamiltonian, both in matrix form. We obtain
\
\\

\begin{equation}
    \hat{G}^{r}= \left( 
    \begin{matrix}
        \hat{g}^{-1}_{MA}      & \hat{\mathcal{H}}^{\dagger}_{MA} & \hat{0}                          & \hat{0} \\
        \hat{\mathcal{H}}_{MA} & \hat{g}^{-1}_{1}                 & \hat{\Lambda}                    & \hat{0} \\
        \hat{0}                & -\hat{\Lambda}^{*}               & \hat{g}^{-1}_{2}                 & \hat{\mathcal{H}}_{MB} \\
        \hat{0}                & \hat{0}                          & \hat{\mathcal{H}}^{\dagger}_{MB} & \hat{g}^{-1}_{MB} 
    \end{matrix}
    \right)^{-1}\,,
    \label{Gr1}
\end{equation}
where each element in Eq. (\ref{Gr1}) corresponds to a 2 x 2 matrix, and are given by
\begin{equation}
    \hat{g}^{-1}_{MA(B)} = \left( 
    \begin{matrix}
        \omega-\varepsilon_{_{MA(B)}} & 0  \\
        0                             & \omega+\varepsilon_{_{MA(B)}}
    \end{matrix} 
    \right)\,,
    \label{Gr2}
\end{equation}
for MZMs, and
\begin{equation}
    \hat{g}^{-1}_{1(2)}= \left( 
    \begin{matrix}
        \omega-\varepsilon_{_{1(2)}}+\dfrac{i}{2}\sum\limits_{\alpha}\Gamma^{\alpha}_{11(22)}   &   0  \\
        0   &   \omega+\varepsilon_{_{1(2)}}+\dfrac{i}{2}\sum\limits_{\alpha}\Gamma^{\alpha}_{11(22)}
    \end{matrix} 
    \right)\,,
    \label{Gr3}
\end{equation}
for QDs, in which the leads contribution are considered. The QD-MZM coupling matrix is 
\begin{equation}
    \hat{H}_{MA(B)} = \dfrac{1}{\sqrt{2}}\left( 
    \begin{matrix}
        \lambda^{*}_{A(B)} & \lambda^{*}_{A(B)}  \\
        -\lambda_{A(B)}    & -\lambda_{A(B)}
    \end{matrix} \right)\,,
    \label{Gr4}
\end{equation}
while the inter-QDs coupling matrix is
\begin{equation}
    \hat{\Lambda}= \left(
    \begin{matrix}
        \dfrac{i}{2}\sum\limits_{\alpha}\Lambda^{\alpha}_{12}   &   0  \\
        0   &   \dfrac{i}{2}\sum\limits_{\alpha}\Lambda^{\alpha}_{12}
    \end{matrix} 
    \right)\,.
    \label{Gr5}
\end{equation}

Obtaining the inversion of Eq.\,(\ref{Gr1}), one can identify the necessary matrix elements for the quantities under study. For the LDOS of each QD, the corresponding Green's function is located according to the first diagonal element of Eq.\,(\ref{Gr3}). Besides, as the MZM operators are described as a superposition of regular fermionic operators Eqs.\,(\ref{7})-(\ref{8}), the Green's function for $\gamma_{_{j,\beta}}$, $G^{r}_{\eta, \beta}(\omega)$, is obtained from the addition of all elements located according to Eq.\,(\ref{Gr2}).

%%%%%%%%%%%%%%%%%%%%%%%%%%%%%%%%%%%%%%%%%%%%%%%%%%%%%%%%%%%%%%
\section{Full Green's function poles}
%%%%%%%%%%%%%%%%%%%%%%%%%%%%%%%%%%%%%%%%%%%%%%%%%%%%%%%%%%%%%%

We obtain the full Green's function poles from Eq.\,(\ref{Gr1}). For the particular case of using $\lambda_{A}=\lambda_{B}=\lambda$, $\varepsilon_{_{MA}}=\varepsilon_{_{MB}}=\varepsilon_{_{M}}$ and  $\varepsilon_{_{1}}=\varepsilon_{_{2}}=\varepsilon$, the poles are described by the roots of

\begin{equation}
    P_0 + P_1\omega + P_2\omega^2 + P_3\omega^3 + P_4\omega^4 + P_5\omega^5 + P_6\omega^6 + P_7\omega^7 + \omega^8 = 0\,,
\end{equation}
where the coefficients $P_i$ are defined as follows
\begin{eqnarray}
      P_0 = & & \varepsilon_{_{M}}^4\left[ \Gamma^4 \sin^{4}(\phi/2) + 2\varepsilon^2\Gamma^2\left( \cos^{2}(\phi/2) + 1 \right) + \varepsilon^2 \right]\,,                         \\
      P_1 = &-& 4i\Gamma\varepsilon_{_{M}}^2\left[ \varepsilon_{_{M}}^2 + \lambda^2 \right] \left[ \varepsilon^2 + \Gamma^2 \sin^{2}(\phi/2)  \right]\,,                            \\
      P_2 = &-& 2\varepsilon_{_{M}}^2\Gamma^2\left[ \left( 2\lambda^2 + \varepsilon_{_{M}}^2 \right)\left(3 - \cos^{2}(\phi/2) \right) + \Gamma^{2}\sin^{4}(\phi/2)\right]          % \nonumber\\
            -\varepsilon^2\varepsilon_{_{M}}^2\left[ 2\Gamma^2\left( \cos^{2}(\phi/2) + 1 \right) + \varepsilon^2 + 2\lambda^2 + \varepsilon_{_{M}}^2  \right]                      \nonumber\\
            &-& 4\lambda^{4}\Gamma^2 \sin^{2}(\phi/2)\,,                                                                                                                            \\
      P_3 = & & 4i\Gamma\left[ \left(\varepsilon_{_{M}}^2 + \lambda^{2}\right)\left( \Gamma^2\sin^{2}(\phi/2) + \varepsilon^2 + 2\lambda^2 + \varepsilon_{_{M}}^2 \right) \right.   % \nonumber\\
            + \left. \varepsilon_{_{M}}^2\left( \varepsilon^2 + \Gamma^2\sin^{2}(\phi/2) \right) \right]\,,                                                                         \\
      P_4 = & & 4\left[\lambda^2 + \varepsilon_{_{M}}^{2}\right] \left[ \lambda^2 + \varepsilon^{2} + \Gamma^2\left(3 - \cos^{2}(\phi/2) \right)\right]  + \varepsilon_{_{M}}^{4}   % \nonumber\\
            + \varepsilon^2\left[ \varepsilon^2 + 2\Gamma^2\left(\cos^{2}(\phi/2) + 1 \right)  \right] + \Gamma^4\sin^{4}(\phi/2)\,,                                                \\
      P_5 = &-& 4i\Gamma\left[ \Gamma^2 \sin^{2}(\phi/2) + \varepsilon^2 + 3\lambda^2 + 2\varepsilon_{_{M}}^2 \right]\,,                                                            \\
      P_6 = &-& 2\left[ \Gamma^2 \left(3 - \cos^{2}(\phi/2) \right) + \varepsilon^2 + 2\lambda^2 + \varepsilon_{_{M}}^2 \right]\,,                                                  \\
      P_7 = & & 4i\Gamma\,.                                                                                                                                                                           
\end{eqnarray}

\twocolumngrid

%\nocite{*}

\bibliographystyle{apsrev4-1}
\bibliography{bibprepint}% Produces the bibliography via BibTeX.

%merlin.mbs apsrev4-1.bst 2010-07-25 4.21a (PWD, AO, DPC) hacked
%Control: key (0)
%Control: author (72) initials jnrlst
%Control: editor formatted (1) identically to author
%Control: production of article title (-1) disabled
%Control: page (0) single
%Control: year (1) truncated
%Control: production of eprint (0) enabled
\providecommand{\noopsort}[1]{}\providecommand{\singleletter}[1]{#1}%
\begin{thebibliography}{69}%
\makeatletter
\providecommand \@ifxundefined [1]{%
 \@ifx{#1\undefined}
}%
\providecommand \@ifnum [1]{%
 \ifnum #1\expandafter \@firstoftwo
 \else \expandafter \@secondoftwo
 \fi
}%
\providecommand \@ifx [1]{%
 \ifx #1\expandafter \@firstoftwo
 \else \expandafter \@secondoftwo
 \fi
}%
\providecommand \natexlab [1]{#1}%
\providecommand \enquote  [1]{``#1''}%
\providecommand \bibnamefont  [1]{#1}%
\providecommand \bibfnamefont [1]{#1}%
\providecommand \citenamefont [1]{#1}%
\providecommand \href@noop [0]{\@secondoftwo}%
\providecommand \href [0]{\begingroup \@sanitize@url \@href}%
\providecommand \@href[1]{\@@startlink{#1}\@@href}%
\providecommand \@@href[1]{\endgroup#1\@@endlink}%
\providecommand \@sanitize@url [0]{\catcode `\\12\catcode `\$12\catcode
  `\&12\catcode `\#12\catcode `\^12\catcode `\_12\catcode `\%12\relax}%
\providecommand \@@startlink[1]{}%
\providecommand \@@endlink[0]{}%
\providecommand \url  [0]{\begingroup\@sanitize@url \@url }%
\providecommand \@url [1]{\endgroup\@href {#1}{\urlprefix }}%
\providecommand \urlprefix  [0]{URL }%
\providecommand \Eprint [0]{\href }%
\providecommand \doibase [0]{http://dx.doi.org/}%
\providecommand \selectlanguage [0]{\@gobble}%
\providecommand \bibinfo  [0]{\@secondoftwo}%
\providecommand \bibfield  [0]{\@secondoftwo}%
\providecommand \translation [1]{[#1]}%
\providecommand \BibitemOpen [0]{}%
\providecommand \bibitemStop [0]{}%
\providecommand \bibitemNoStop [0]{.\EOS\space}%
\providecommand \EOS [0]{\spacefactor3000\relax}%
\providecommand \BibitemShut  [1]{\csname bibitem#1\endcsname}%
\let\auto@bib@innerbib\@empty
%</preamble>
\bibitem [{\citenamefont {Kitaev}(2003)}]{kitaev2003fault}%
  \BibitemOpen
  \bibfield  {author} {\bibinfo {author} {\bibfnamefont {A.~Y.}\ \bibnamefont
  {Kitaev}},\ }\href
  {https://www.sciencedirect.com/science/article/pii/S0003491602000180}
  {\bibfield  {journal} {\bibinfo  {journal} {Ann. Phys.}\ }\textbf {\bibinfo
  {volume} {303}},\ \bibinfo {pages} {2} (\bibinfo {year} {2003})}\BibitemShut
  {NoStop}%
\bibitem [{\citenamefont {Nayak}\ \emph {et~al.}(2008)\citenamefont {Nayak},
  \citenamefont {Simon}, \citenamefont {Stern}, \citenamefont {Freedman},\ and\
  \citenamefont {Sarma}}]{nayak2008non}%
  \BibitemOpen
  \bibfield  {author} {\bibinfo {author} {\bibfnamefont {C.}~\bibnamefont
  {Nayak}}, \bibinfo {author} {\bibfnamefont {S.~H.}\ \bibnamefont {Simon}},
  \bibinfo {author} {\bibfnamefont {A.}~\bibnamefont {Stern}}, \bibinfo
  {author} {\bibfnamefont {M.}~\bibnamefont {Freedman}}, \ and\ \bibinfo
  {author} {\bibfnamefont {S.~D.}\ \bibnamefont {Sarma}},\ }\href
  {https://journals.aps.org/rmp/abstract/10.1103/RevModPhys.80.1083} {\bibfield
   {journal} {\bibinfo  {journal} {Rev. Mod. Phys.}\ }\textbf {\bibinfo
  {volume} {80}},\ \bibinfo {pages} {1083} (\bibinfo {year}
  {2008})}\BibitemShut {NoStop}%
\bibitem [{\citenamefont {Pachos}(2012)}]{pachos2012introduction}%
  \BibitemOpen
  \bibfield  {author} {\bibinfo {author} {\bibfnamefont {J.~K.}\ \bibnamefont
  {Pachos}},\ }\href
  {https://www.cambridge.org/cl/academic/subjects/physics/quantum-physics-quantum-information-and-quantum-computation/introduction-topological-quantum-computation?format=HB&isbn=9781107005044}
  {\emph {\bibinfo {title} {Introduction to topological quantum computation}}}\
  (\bibinfo  {publisher} {Cambridge University Press},\ \bibinfo {year}
  {2012})\BibitemShut {NoStop}%
\bibitem [{\citenamefont {Beenakker}(2013)}]{beenakker2013search}%
  \BibitemOpen
  \bibfield  {author} {\bibinfo {author} {\bibfnamefont {C.~W.~J.}\
  \bibnamefont {Beenakker}},\ }\href
  {https://www.annualreviews.org/doi/abs/10.1146/annurev-conmatphys-030212-184337}
  {\bibfield  {journal} {\bibinfo  {journal} {Annu. Rev. Condens. Matter
  Phys.}\ }\textbf {\bibinfo {volume} {4}},\ \bibinfo {pages} {113} (\bibinfo
  {year} {2013})}\BibitemShut {NoStop}%
\bibitem [{\citenamefont {Laflamme}\ \emph {et~al.}(2014)\citenamefont
  {Laflamme}, \citenamefont {Baranov}, \citenamefont {Zoller},\ and\
  \citenamefont {Kraus}}]{laflamme2014publisher}%
  \BibitemOpen
  \bibfield  {author} {\bibinfo {author} {\bibfnamefont {C.}~\bibnamefont
  {Laflamme}}, \bibinfo {author} {\bibfnamefont {M.}~\bibnamefont {Baranov}},
  \bibinfo {author} {\bibfnamefont {P.}~\bibnamefont {Zoller}}, \ and\ \bibinfo
  {author} {\bibfnamefont {C.}~\bibnamefont {Kraus}},\ }\href
  {https://journals.aps.org/pra/abstract/10.1103/PhysRevA.89.022319} {\bibfield
   {journal} {\bibinfo  {journal} {Phys. Rev. A}\ }\textbf {\bibinfo {volume}
  {89}},\ \bibinfo {pages} {029903} (\bibinfo {year} {2014})}\BibitemShut
  {NoStop}%
\bibitem [{\citenamefont {Albrecht}\ \emph {et~al.}(2016)\citenamefont
  {Albrecht}, \citenamefont {Higginbotham}, \citenamefont {Madsen},
  \citenamefont {Kuemmeth}, \citenamefont {Jespersen}, \citenamefont
  {Nyg{\aa}rd}, \citenamefont {Krogstrup},\ and\ \citenamefont
  {Marcus}}]{albrecht2016exponential}%
  \BibitemOpen
  \bibfield  {author} {\bibinfo {author} {\bibfnamefont {S.~M.}\ \bibnamefont
  {Albrecht}}, \bibinfo {author} {\bibfnamefont {A.~P.}\ \bibnamefont
  {Higginbotham}}, \bibinfo {author} {\bibfnamefont {M.}~\bibnamefont
  {Madsen}}, \bibinfo {author} {\bibfnamefont {F.}~\bibnamefont {Kuemmeth}},
  \bibinfo {author} {\bibfnamefont {T.~S.}\ \bibnamefont {Jespersen}}, \bibinfo
  {author} {\bibfnamefont {J.}~\bibnamefont {Nyg{\aa}rd}}, \bibinfo {author}
  {\bibfnamefont {P.}~\bibnamefont {Krogstrup}}, \ and\ \bibinfo {author}
  {\bibfnamefont {C.}~\bibnamefont {Marcus}},\ }\href
  {https://www.nature.com/articles/nature17162} {\bibfield  {journal} {\bibinfo
   {journal} {Nature}\ }\textbf {\bibinfo {volume} {531}},\ \bibinfo {pages}
  {206} (\bibinfo {year} {2016})}\BibitemShut {NoStop}%
\bibitem [{\citenamefont {Majorana}(1937)}]{majorana1937teoria}%
  \BibitemOpen
  \bibfield  {author} {\bibinfo {author} {\bibfnamefont {E.}~\bibnamefont
  {Majorana}},\ }\href {https://link.springer.com/article/10.1007/BF02961314}
  {\bibfield  {journal} {\bibinfo  {journal} {Nuovo Cimento}\ }\textbf
  {\bibinfo {volume} {14}},\ \bibinfo {pages} {171} (\bibinfo {year}
  {1937})}\BibitemShut {NoStop}%
\bibitem [{\citenamefont {Wilczek}(2009)}]{wilczek2009majorana}%
  \BibitemOpen
  \bibfield  {author} {\bibinfo {author} {\bibfnamefont {F.}~\bibnamefont
  {Wilczek}},\ }\href {https://www.nature.com/articles/nphys1380} {\bibfield
  {journal} {\bibinfo  {journal} {Nat. Phys.}\ }\textbf {\bibinfo {volume}
  {5}},\ \bibinfo {pages} {614} (\bibinfo {year} {2009})}\BibitemShut {NoStop}%
\bibitem [{\citenamefont {Franz}(2010)}]{franz2010race}%
  \BibitemOpen
  \bibfield  {author} {\bibinfo {author} {\bibfnamefont {M.}~\bibnamefont
  {Franz}},\ }\href {https://physics.aps.org/articles/v3/24} {\bibfield
  {journal} {\bibinfo  {journal} {Physics}\ }\textbf {\bibinfo {volume} {3}},\
  \bibinfo {pages} {24} (\bibinfo {year} {2010})}\BibitemShut {NoStop}%
\bibitem [{\citenamefont {Kraus}\ \emph {et~al.}(2013)\citenamefont {Kraus},
  \citenamefont {Dalmonte}, \citenamefont {Baranov}, \citenamefont
  {L{\"a}uchli},\ and\ \citenamefont {Zoller}}]{kraus2013majorana}%
  \BibitemOpen
  \bibfield  {author} {\bibinfo {author} {\bibfnamefont {C.~V.}\ \bibnamefont
  {Kraus}}, \bibinfo {author} {\bibfnamefont {M.}~\bibnamefont {Dalmonte}},
  \bibinfo {author} {\bibfnamefont {M.~A.}\ \bibnamefont {Baranov}}, \bibinfo
  {author} {\bibfnamefont {A.~M.}\ \bibnamefont {L{\"a}uchli}}, \ and\ \bibinfo
  {author} {\bibfnamefont {P.}~\bibnamefont {Zoller}},\ }\href
  {https://journals.aps.org/prl/abstract/10.1103/PhysRevLett.111.173004}
  {\bibfield  {journal} {\bibinfo  {journal} {Phys. Rev. Lett.}\ }\textbf
  {\bibinfo {volume} {111}},\ \bibinfo {pages} {173004} (\bibinfo {year}
  {2013})}\BibitemShut {NoStop}%
\bibitem [{\citenamefont {Alicea}\ \emph {et~al.}(2011)\citenamefont {Alicea},
  \citenamefont {Oreg}, \citenamefont {Refael}, \citenamefont {Von~Oppen},\
  and\ \citenamefont {Fisher}}]{alicea2011non}%
  \BibitemOpen
  \bibfield  {author} {\bibinfo {author} {\bibfnamefont {J.}~\bibnamefont
  {Alicea}}, \bibinfo {author} {\bibfnamefont {Y.}~\bibnamefont {Oreg}},
  \bibinfo {author} {\bibfnamefont {G.}~\bibnamefont {Refael}}, \bibinfo
  {author} {\bibfnamefont {F.}~\bibnamefont {Von~Oppen}}, \ and\ \bibinfo
  {author} {\bibfnamefont {M.}~\bibnamefont {Fisher}},\ }\href
  {https://www.nature.com/articles/nphys1915} {\bibfield  {journal} {\bibinfo
  {journal} {Nat. Phys.}\ }\textbf {\bibinfo {volume} {7}},\ \bibinfo {pages}
  {412} (\bibinfo {year} {2011})}\BibitemShut {NoStop}%
\bibitem [{\citenamefont {Kitaev}(2001)}]{kitaev2001unpaired}%
  \BibitemOpen
  \bibfield  {author} {\bibinfo {author} {\bibfnamefont {A.~Y.}\ \bibnamefont
  {Kitaev}},\ }\href
  {https://iopscience.iop.org/article/10.1070/1063-7869/44/10S/S29/meta}
  {\bibfield  {journal} {\bibinfo  {journal} {Phys.-Usp.}\ }\textbf {\bibinfo
  {volume} {44}},\ \bibinfo {pages} {131} (\bibinfo {year} {2001})}\BibitemShut
  {NoStop}%
\bibitem [{\citenamefont {Bravyi}\ and\ \citenamefont
  {Kitaev}(2002)}]{bravyi2002fermionic}%
  \BibitemOpen
  \bibfield  {author} {\bibinfo {author} {\bibfnamefont {S.~B.}\ \bibnamefont
  {Bravyi}}\ and\ \bibinfo {author} {\bibfnamefont {A.~Y.}\ \bibnamefont
  {Kitaev}},\ }\href
  {https://www.sciencedirect.com/science/article/pii/S0003491602962548}
  {\bibfield  {journal} {\bibinfo  {journal} {Ann. Phys.}\ }\textbf {\bibinfo
  {volume} {298}},\ \bibinfo {pages} {210} (\bibinfo {year}
  {2002})}\BibitemShut {NoStop}%
\bibitem [{\citenamefont {Leijnse}\ and\ \citenamefont
  {Flensberg}(2011)}]{leijnse2011quantum}%
  \BibitemOpen
  \bibfield  {author} {\bibinfo {author} {\bibfnamefont {M.}~\bibnamefont
  {Leijnse}}\ and\ \bibinfo {author} {\bibfnamefont {K.}~\bibnamefont
  {Flensberg}},\ }\href
  {https://journals.aps.org/prl/abstract/10.1103/PhysRevLett.107.210502}
  {\bibfield  {journal} {\bibinfo  {journal} {Phys. Rev. Lett.}\ }\textbf
  {\bibinfo {volume} {107}},\ \bibinfo {pages} {210502} (\bibinfo {year}
  {2011})}\BibitemShut {NoStop}%
\bibitem [{\citenamefont {Moore}(2009)}]{moore2009next}%
  \BibitemOpen
  \bibfield  {author} {\bibinfo {author} {\bibfnamefont {J.}~\bibnamefont
  {Moore}},\ }\href {https://www.nature.com/articles/nphys1294} {\bibfield
  {journal} {\bibinfo  {journal} {Nat. Phys.}\ }\textbf {\bibinfo {volume}
  {5}},\ \bibinfo {pages} {378} (\bibinfo {year} {2009})}\BibitemShut {NoStop}%
\bibitem [{\citenamefont {Wu}\ and\ \citenamefont
  {Cao}(2012)}]{wu2012tunneling}%
  \BibitemOpen
  \bibfield  {author} {\bibinfo {author} {\bibfnamefont {B.}~\bibnamefont
  {Wu}}\ and\ \bibinfo {author} {\bibfnamefont {J.}~\bibnamefont {Cao}},\
  }\href {https://journals.aps.org/prb/abstract/10.1103/PhysRevB.85.085415}
  {\bibfield  {journal} {\bibinfo  {journal} {Phys. Rev. B}\ }\textbf {\bibinfo
  {volume} {85}},\ \bibinfo {pages} {085415} (\bibinfo {year}
  {2012})}\BibitemShut {NoStop}%
\bibitem [{\citenamefont {Semenoff}\ and\ \citenamefont
  {Sodano}(2006)}]{semenoff2006teleportation}%
  \BibitemOpen
  \bibfield  {author} {\bibinfo {author} {\bibfnamefont {G.~W.}\ \bibnamefont
  {Semenoff}}\ and\ \bibinfo {author} {\bibfnamefont {P.}~\bibnamefont
  {Sodano}},\ }\href {https://arxiv.org/abs/cond-mat/0601261} {\bibfield
  {journal} {\bibinfo  {journal} {arXiv:cond-mat/0601261}\ } (\bibinfo {year}
  {2006})}\BibitemShut {NoStop}%
\bibitem [{\citenamefont {Tewari}\ \emph {et~al.}(2008)\citenamefont {Tewari},
  \citenamefont {Zhang}, \citenamefont {Sarma}, \citenamefont {Nayak},\ and\
  \citenamefont {Lee}}]{tewari2008testable}%
  \BibitemOpen
  \bibfield  {author} {\bibinfo {author} {\bibfnamefont {S.}~\bibnamefont
  {Tewari}}, \bibinfo {author} {\bibfnamefont {C.}~\bibnamefont {Zhang}},
  \bibinfo {author} {\bibfnamefont {S.~D.}\ \bibnamefont {Sarma}}, \bibinfo
  {author} {\bibfnamefont {C.}~\bibnamefont {Nayak}}, \ and\ \bibinfo {author}
  {\bibfnamefont {D.-H.}\ \bibnamefont {Lee}},\ }\href
  {https://journals.aps.org/prl/abstract/10.1103/PhysRevLett.100.027001}
  {\bibfield  {journal} {\bibinfo  {journal} {Phys. Rev. lett.}\ }\textbf
  {\bibinfo {volume} {100}},\ \bibinfo {pages} {027001} (\bibinfo {year}
  {2008})}\BibitemShut {NoStop}%
\bibitem [{\citenamefont {Mourik}\ \emph {et~al.}(2012)\citenamefont {Mourik},
  \citenamefont {Zuo}, \citenamefont {Frolov}, \citenamefont {Plissard},
  \citenamefont {Bakkers},\ and\ \citenamefont
  {Kouwenhoven}}]{mourik2012signatures}%
  \BibitemOpen
  \bibfield  {author} {\bibinfo {author} {\bibfnamefont {V.}~\bibnamefont
  {Mourik}}, \bibinfo {author} {\bibfnamefont {K.}~\bibnamefont {Zuo}},
  \bibinfo {author} {\bibfnamefont {S.~M.}\ \bibnamefont {Frolov}}, \bibinfo
  {author} {\bibfnamefont {S.}~\bibnamefont {Plissard}}, \bibinfo {author}
  {\bibfnamefont {E.~P.}\ \bibnamefont {Bakkers}}, \ and\ \bibinfo {author}
  {\bibfnamefont {L.~P.}\ \bibnamefont {Kouwenhoven}},\ }\href
  {https://www.science.org/doi/full/10.1126/science.1222360} {\bibfield
  {journal} {\bibinfo  {journal} {Science}\ }\textbf {\bibinfo {volume}
  {336}},\ \bibinfo {pages} {1003} (\bibinfo {year} {2012})}\BibitemShut
  {NoStop}%
\bibitem [{\citenamefont {Bolech}\ and\ \citenamefont
  {Demler}(2007)}]{bolech2007observing}%
  \BibitemOpen
  \bibfield  {author} {\bibinfo {author} {\bibfnamefont {C.}~\bibnamefont
  {Bolech}}\ and\ \bibinfo {author} {\bibfnamefont {E.}~\bibnamefont
  {Demler}},\ }\href
  {https://journals.aps.org/prl/abstract/10.1103/PhysRevLett.98.237002}
  {\bibfield  {journal} {\bibinfo  {journal} {Phys. Rev. Lett.}\ }\textbf
  {\bibinfo {volume} {98}},\ \bibinfo {pages} {237002} (\bibinfo {year}
  {2007})}\BibitemShut {NoStop}%
\bibitem [{\citenamefont {Nilsson}\ \emph {et~al.}(2008)\citenamefont
  {Nilsson}, \citenamefont {Akhmerov},\ and\ \citenamefont
  {Beenakker}}]{nilsson2008splitting}%
  \BibitemOpen
  \bibfield  {author} {\bibinfo {author} {\bibfnamefont {J.}~\bibnamefont
  {Nilsson}}, \bibinfo {author} {\bibfnamefont {A.}~\bibnamefont {Akhmerov}}, \
  and\ \bibinfo {author} {\bibfnamefont {C.~W.~J.}\ \bibnamefont {Beenakker}},\
  }\href {https://journals.aps.org/prl/abstract/10.1103/PhysRevLett.101.120403}
  {\bibfield  {journal} {\bibinfo  {journal} {Phys. Rev. Lett.}\ }\textbf
  {\bibinfo {volume} {101}},\ \bibinfo {pages} {120403} (\bibinfo {year}
  {2008})}\BibitemShut {NoStop}%
\bibitem [{\citenamefont {Law}\ \emph {et~al.}(2009)\citenamefont {Law},
  \citenamefont {Lee},\ and\ \citenamefont {Ng}}]{law2009majorana}%
  \BibitemOpen
  \bibfield  {author} {\bibinfo {author} {\bibfnamefont {K.~T.}\ \bibnamefont
  {Law}}, \bibinfo {author} {\bibfnamefont {P.~A.}\ \bibnamefont {Lee}}, \ and\
  \bibinfo {author} {\bibfnamefont {T.~K.}\ \bibnamefont {Ng}},\ }\href
  {https://journals.aps.org/prl/abstract/10.1103/PhysRevLett.103.237001}
  {\bibfield  {journal} {\bibinfo  {journal} {Phys. Rev. Lett.}\ }\textbf
  {\bibinfo {volume} {103}},\ \bibinfo {pages} {237001} (\bibinfo {year}
  {2009})}\BibitemShut {NoStop}%
\bibitem [{\citenamefont {Fu}\ and\ \citenamefont
  {Kane}(2009)}]{fu2009josephson}%
  \BibitemOpen
  \bibfield  {author} {\bibinfo {author} {\bibfnamefont {L.}~\bibnamefont
  {Fu}}\ and\ \bibinfo {author} {\bibfnamefont {C.~L.}\ \bibnamefont {Kane}},\
  }\href {https://journals.aps.org/prb/abstract/10.1103/PhysRevB.79.161408}
  {\bibfield  {journal} {\bibinfo  {journal} {Phys. Rev. B}\ }\textbf {\bibinfo
  {volume} {79}},\ \bibinfo {pages} {161408(R)} (\bibinfo {year}
  {2009})}\BibitemShut {NoStop}%
\bibitem [{\citenamefont {Flensberg}(2010)}]{flensberg2010tunneling}%
  \BibitemOpen
  \bibfield  {author} {\bibinfo {author} {\bibfnamefont {K.}~\bibnamefont
  {Flensberg}},\ }\href
  {https://journals.aps.org/prb/abstract/10.1103/PhysRevB.82.180516} {\bibfield
   {journal} {\bibinfo  {journal} {Phys. Rev. B}\ }\textbf {\bibinfo {volume}
  {82}},\ \bibinfo {pages} {180516(R)} (\bibinfo {year} {2010})}\BibitemShut
  {NoStop}%
\bibitem [{\citenamefont {Pikulin}\ \emph {et~al.}(2012)\citenamefont
  {Pikulin}, \citenamefont {Dahlhaus}, \citenamefont {Wimmer}, \citenamefont
  {Schomerus},\ and\ \citenamefont {Beenakker}}]{pikulin2012h}%
  \BibitemOpen
  \bibfield  {author} {\bibinfo {author} {\bibfnamefont {D.}~\bibnamefont
  {Pikulin}}, \bibinfo {author} {\bibfnamefont {J.}~\bibnamefont {Dahlhaus}},
  \bibinfo {author} {\bibfnamefont {M.}~\bibnamefont {Wimmer}}, \bibinfo
  {author} {\bibfnamefont {H.}~\bibnamefont {Schomerus}}, \ and\ \bibinfo
  {author} {\bibfnamefont {C.~W.~J.}\ \bibnamefont {Beenakker}},\ }\href
  {https://iopscience.iop.org/article/10.1088/1367-2630/14/12/125011/meta}
  {\bibfield  {journal} {\bibinfo  {journal} {New J. Phys.}\ }\textbf {\bibinfo
  {volume} {14}},\ \bibinfo {pages} {125011} (\bibinfo {year}
  {2012})}\BibitemShut {NoStop}%
\bibitem [{\citenamefont {Franz}(2013)}]{franz2013majorana}%
  \BibitemOpen
  \bibfield  {author} {\bibinfo {author} {\bibfnamefont {M.}~\bibnamefont
  {Franz}},\ }\href {https://www.nature.com/articles/nnano.2013.33} {\bibfield
  {journal} {\bibinfo  {journal} {Nat. Nanotechnol.}\ }\textbf {\bibinfo
  {volume} {8}},\ \bibinfo {pages} {149} (\bibinfo {year} {2013})}\BibitemShut
  {NoStop}%
\bibitem [{\citenamefont {Prada}\ \emph {et~al.}(2012)\citenamefont {Prada},
  \citenamefont {San-Jose},\ and\ \citenamefont {Aguado}}]{prada2012transport}%
  \BibitemOpen
  \bibfield  {author} {\bibinfo {author} {\bibfnamefont {E.}~\bibnamefont
  {Prada}}, \bibinfo {author} {\bibfnamefont {P.}~\bibnamefont {San-Jose}}, \
  and\ \bibinfo {author} {\bibfnamefont {R.}~\bibnamefont {Aguado}},\ }\href
  {https://journals.aps.org/prb/abstract/10.1103/PhysRevB.86.180503} {\bibfield
   {journal} {\bibinfo  {journal} {Phys. Rev. B}\ }\textbf {\bibinfo {volume}
  {86}},\ \bibinfo {pages} {180503(R)} (\bibinfo {year} {2012})}\BibitemShut
  {NoStop}%
\bibitem [{\citenamefont {Rainis}\ \emph {et~al.}(2013)\citenamefont {Rainis},
  \citenamefont {Trifunovic}, \citenamefont {Klinovaja},\ and\ \citenamefont
  {Loss}}]{rainis2013towards}%
  \BibitemOpen
  \bibfield  {author} {\bibinfo {author} {\bibfnamefont {D.}~\bibnamefont
  {Rainis}}, \bibinfo {author} {\bibfnamefont {L.}~\bibnamefont {Trifunovic}},
  \bibinfo {author} {\bibfnamefont {J.}~\bibnamefont {Klinovaja}}, \ and\
  \bibinfo {author} {\bibfnamefont {D.}~\bibnamefont {Loss}},\ }\href
  {https://journals.aps.org/prb/abstract/10.1103/PhysRevB.87.024515} {\bibfield
   {journal} {\bibinfo  {journal} {Phys. Rev. B}\ }\textbf {\bibinfo {volume}
  {87}},\ \bibinfo {pages} {024515} (\bibinfo {year} {2013})}\BibitemShut
  {NoStop}%
\bibitem [{\citenamefont {Cook}\ \emph {et~al.}(2012)\citenamefont {Cook},
  \citenamefont {Vazifeh},\ and\ \citenamefont {Franz}}]{cook2012stability}%
  \BibitemOpen
  \bibfield  {author} {\bibinfo {author} {\bibfnamefont {A.}~\bibnamefont
  {Cook}}, \bibinfo {author} {\bibfnamefont {M.}~\bibnamefont {Vazifeh}}, \
  and\ \bibinfo {author} {\bibfnamefont {M.}~\bibnamefont {Franz}},\ }\href
  {https://journals.aps.org/prb/abstract/10.1103/PhysRevB.86.155431} {\bibfield
   {journal} {\bibinfo  {journal} {Phys. Rev. B}\ }\textbf {\bibinfo {volume}
  {86}},\ \bibinfo {pages} {155431} (\bibinfo {year} {2012})}\BibitemShut
  {NoStop}%
\bibitem [{\citenamefont {Liu}\ and\ \citenamefont
  {Lobos}(2013)}]{liu2013manipulating}%
  \BibitemOpen
  \bibfield  {author} {\bibinfo {author} {\bibfnamefont {X.-J.}\ \bibnamefont
  {Liu}}\ and\ \bibinfo {author} {\bibfnamefont {A.~M.}\ \bibnamefont
  {Lobos}},\ }\href
  {https://journals.aps.org/prb/abstract/10.1103/PhysRevB.87.060504} {\bibfield
   {journal} {\bibinfo  {journal} {Phys. Rev. B}\ }\textbf {\bibinfo {volume}
  {87}},\ \bibinfo {pages} {060504} (\bibinfo {year} {2013})}\BibitemShut
  {NoStop}%
\bibitem [{\citenamefont {Stanescu}\ \emph {et~al.}(2011)\citenamefont
  {Stanescu}, \citenamefont {Lutchyn},\ and\ \citenamefont
  {Sarma}}]{stanescu2011majorana}%
  \BibitemOpen
  \bibfield  {author} {\bibinfo {author} {\bibfnamefont {T.~D.}\ \bibnamefont
  {Stanescu}}, \bibinfo {author} {\bibfnamefont {R.~M.}\ \bibnamefont
  {Lutchyn}}, \ and\ \bibinfo {author} {\bibfnamefont {S.~D.}\ \bibnamefont
  {Sarma}},\ }\href
  {https://journals.aps.org/prb/abstract/10.1103/PhysRevB.84.144522} {\bibfield
   {journal} {\bibinfo  {journal} {Phys. Rev. B}\ }\textbf {\bibinfo {volume}
  {84}},\ \bibinfo {pages} {144522} (\bibinfo {year} {2011})}\BibitemShut
  {NoStop}%
\bibitem [{\citenamefont {Lee}\ \emph {et~al.}(2014)\citenamefont {Lee},
  \citenamefont {Jiang}, \citenamefont {Houzet}, \citenamefont {Aguado},
  \citenamefont {Lieber},\ and\ \citenamefont {De~Franceschi}}]{lee2014spin}%
  \BibitemOpen
  \bibfield  {author} {\bibinfo {author} {\bibfnamefont {E.~J.~H.}\
  \bibnamefont {Lee}}, \bibinfo {author} {\bibfnamefont {X.}~\bibnamefont
  {Jiang}}, \bibinfo {author} {\bibfnamefont {M.}~\bibnamefont {Houzet}},
  \bibinfo {author} {\bibfnamefont {R.}~\bibnamefont {Aguado}}, \bibinfo
  {author} {\bibfnamefont {C.~M.}\ \bibnamefont {Lieber}}, \ and\ \bibinfo
  {author} {\bibfnamefont {S.}~\bibnamefont {De~Franceschi}},\ }\href
  {https://www.nature.com/articles/nnano.2013.267} {\bibfield  {journal}
  {\bibinfo  {journal} {Nat. Nanotechnol.}\ }\textbf {\bibinfo {volume} {9}},\
  \bibinfo {pages} {79} (\bibinfo {year} {2014})}\BibitemShut {NoStop}%
\bibitem [{\citenamefont {Wimmer}\ \emph {et~al.}(2011)\citenamefont {Wimmer},
  \citenamefont {Akhmerov}, \citenamefont {Dahlhaus},\ and\ \citenamefont
  {Beenakker}}]{wimmer2011quantum}%
  \BibitemOpen
  \bibfield  {author} {\bibinfo {author} {\bibfnamefont {M.}~\bibnamefont
  {Wimmer}}, \bibinfo {author} {\bibfnamefont {A.}~\bibnamefont {Akhmerov}},
  \bibinfo {author} {\bibfnamefont {J.}~\bibnamefont {Dahlhaus}}, \ and\
  \bibinfo {author} {\bibfnamefont {C.}~\bibnamefont {Beenakker}},\ }\href
  {https://iopscience.iop.org/article/10.1088/1367-2630/13/5/053016/meta}
  {\bibfield  {journal} {\bibinfo  {journal} {New J. Phys.}\ }\textbf {\bibinfo
  {volume} {13}},\ \bibinfo {pages} {053016} (\bibinfo {year}
  {2011})}\BibitemShut {NoStop}%
\bibitem [{\citenamefont {Deng}\ \emph {et~al.}(2012)\citenamefont {Deng},
  \citenamefont {Yu}, \citenamefont {Huang}, \citenamefont {Larsson},
  \citenamefont {Caroff},\ and\ \citenamefont {Xu}}]{deng2012anomalous}%
  \BibitemOpen
  \bibfield  {author} {\bibinfo {author} {\bibfnamefont {M.}~\bibnamefont
  {Deng}}, \bibinfo {author} {\bibfnamefont {C.}~\bibnamefont {Yu}}, \bibinfo
  {author} {\bibfnamefont {G.}~\bibnamefont {Huang}}, \bibinfo {author}
  {\bibfnamefont {M.}~\bibnamefont {Larsson}}, \bibinfo {author} {\bibfnamefont
  {P.}~\bibnamefont {Caroff}}, \ and\ \bibinfo {author} {\bibfnamefont
  {H.}~\bibnamefont {Xu}},\ }\href
  {https://pubs.acs.org/doi/full/10.1021/nl303758w} {\bibfield  {journal}
  {\bibinfo  {journal} {Nano lett.}\ }\textbf {\bibinfo {volume} {12}},\
  \bibinfo {pages} {6414} (\bibinfo {year} {2012})}\BibitemShut {NoStop}%
\bibitem [{\citenamefont {Das}\ \emph {et~al.}(2012)\citenamefont {Das},
  \citenamefont {Ronen}, \citenamefont {Most}, \citenamefont {Oreg},
  \citenamefont {Heiblum},\ and\ \citenamefont {Shtrikman}}]{das2012zero}%
  \BibitemOpen
  \bibfield  {author} {\bibinfo {author} {\bibfnamefont {A.}~\bibnamefont
  {Das}}, \bibinfo {author} {\bibfnamefont {Y.}~\bibnamefont {Ronen}}, \bibinfo
  {author} {\bibfnamefont {Y.}~\bibnamefont {Most}}, \bibinfo {author}
  {\bibfnamefont {Y.}~\bibnamefont {Oreg}}, \bibinfo {author} {\bibfnamefont
  {M.}~\bibnamefont {Heiblum}}, \ and\ \bibinfo {author} {\bibfnamefont
  {H.}~\bibnamefont {Shtrikman}},\ }\href
  {https://www.nature.com/articles/nphys2479} {\bibfield  {journal} {\bibinfo
  {journal} {Nat. Phys.}\ }\textbf {\bibinfo {volume} {8}},\ \bibinfo {pages}
  {887} (\bibinfo {year} {2012})}\BibitemShut {NoStop}%
\bibitem [{\citenamefont {Lee}\ \emph {et~al.}(2012)\citenamefont {Lee},
  \citenamefont {Jiang}, \citenamefont {Aguado}, \citenamefont {Katsaros},
  \citenamefont {Lieber},\ and\ \citenamefont {De~Franceschi}}]{lee2012zero}%
  \BibitemOpen
  \bibfield  {author} {\bibinfo {author} {\bibfnamefont {E.~J.~H.}\
  \bibnamefont {Lee}}, \bibinfo {author} {\bibfnamefont {X.}~\bibnamefont
  {Jiang}}, \bibinfo {author} {\bibfnamefont {R.}~\bibnamefont {Aguado}},
  \bibinfo {author} {\bibfnamefont {G.}~\bibnamefont {Katsaros}}, \bibinfo
  {author} {\bibfnamefont {C.~M.}\ \bibnamefont {Lieber}}, \ and\ \bibinfo
  {author} {\bibfnamefont {S.}~\bibnamefont {De~Franceschi}},\ }\href
  {https://journals.aps.org/prl/abstract/10.1103/PhysRevLett.109.186802}
  {\bibfield  {journal} {\bibinfo  {journal} {Phys. Rev. lett.}\ }\textbf
  {\bibinfo {volume} {109}},\ \bibinfo {pages} {186802} (\bibinfo {year}
  {2012})}\BibitemShut {NoStop}%
\bibitem [{\citenamefont {Finck}\ \emph {et~al.}(2013)\citenamefont {Finck},
  \citenamefont {Van~Harlingen}, \citenamefont {Mohseni}, \citenamefont
  {Jung},\ and\ \citenamefont {Li}}]{finck2013anomalous}%
  \BibitemOpen
  \bibfield  {author} {\bibinfo {author} {\bibfnamefont {A.}~\bibnamefont
  {Finck}}, \bibinfo {author} {\bibfnamefont {D.}~\bibnamefont
  {Van~Harlingen}}, \bibinfo {author} {\bibfnamefont {P.}~\bibnamefont
  {Mohseni}}, \bibinfo {author} {\bibfnamefont {K.}~\bibnamefont {Jung}}, \
  and\ \bibinfo {author} {\bibfnamefont {X.}~\bibnamefont {Li}},\ }\href
  {https://journals.aps.org/prl/abstract/10.1103/PhysRevLett.110.126406}
  {\bibfield  {journal} {\bibinfo  {journal} {Phys. Rev. Lett.}\ }\textbf
  {\bibinfo {volume} {110}},\ \bibinfo {pages} {126406} (\bibinfo {year}
  {2013})}\BibitemShut {NoStop}%
\bibitem [{\citenamefont {Churchill}\ \emph {et~al.}(2013)\citenamefont
  {Churchill}, \citenamefont {Fatemi}, \citenamefont {Grove-Rasmussen},
  \citenamefont {Deng}, \citenamefont {Caroff}, \citenamefont {Xu},\ and\
  \citenamefont {Marcus}}]{churchill2013superconductor}%
  \BibitemOpen
  \bibfield  {author} {\bibinfo {author} {\bibfnamefont {H.}~\bibnamefont
  {Churchill}}, \bibinfo {author} {\bibfnamefont {V.}~\bibnamefont {Fatemi}},
  \bibinfo {author} {\bibfnamefont {K.}~\bibnamefont {Grove-Rasmussen}},
  \bibinfo {author} {\bibfnamefont {M.}~\bibnamefont {Deng}}, \bibinfo {author}
  {\bibfnamefont {P.}~\bibnamefont {Caroff}}, \bibinfo {author} {\bibfnamefont
  {H.}~\bibnamefont {Xu}}, \ and\ \bibinfo {author} {\bibfnamefont {C.~M.}\
  \bibnamefont {Marcus}},\ }\href {Superconductor-nanowire devices from
  tunneling to the multichannel regime: Zero-bias oscillations and
  magnetoconductance crossover} {\bibfield  {journal} {\bibinfo  {journal}
  {Phys. Rev. B}\ }\textbf {\bibinfo {volume} {87}},\ \bibinfo {pages} {241401}
  (\bibinfo {year} {2013})}\BibitemShut {NoStop}%
\bibitem [{\citenamefont {Zambrano}\ \emph {et~al.}(2018)\citenamefont
  {Zambrano}, \citenamefont {Ramos-Andrade},\ and\ \citenamefont
  {Orellana}}]{zambrano2018bound}%
  \BibitemOpen
  \bibfield  {author} {\bibinfo {author} {\bibfnamefont {D.}~\bibnamefont
  {Zambrano}}, \bibinfo {author} {\bibfnamefont {J.~P.}\ \bibnamefont
  {Ramos-Andrade}}, \ and\ \bibinfo {author} {\bibfnamefont {P.}~\bibnamefont
  {Orellana}},\ }\href
  {https://iopscience.iop.org/article/10.1088/1361-648X/aad7ca/meta} {\bibfield
   {journal} {\bibinfo  {journal} {J. Phys. Condens. Matter}\ }\textbf
  {\bibinfo {volume} {30}},\ \bibinfo {pages} {375301} (\bibinfo {year}
  {2018})}\BibitemShut {NoStop}%
\bibitem [{\citenamefont {Ramos-Andrade}\ \emph {et~al.}(2019)\citenamefont
  {Ramos-Andrade}, \citenamefont {Zambrano},\ and\ \citenamefont
  {Orellana}}]{ramos2019fano}%
  \BibitemOpen
  \bibfield  {author} {\bibinfo {author} {\bibfnamefont {J.~P.}\ \bibnamefont
  {Ramos-Andrade}}, \bibinfo {author} {\bibfnamefont {D.}~\bibnamefont
  {Zambrano}}, \ and\ \bibinfo {author} {\bibfnamefont {P.~A.}\ \bibnamefont
  {Orellana}},\ }\href
  {https://onlinelibrary.wiley.com/doi/full/10.1002/andp.201800498} {\bibfield
  {journal} {\bibinfo  {journal} {Ann. Phys. (Berlin)}\ }\textbf {\bibinfo
  {volume} {531}},\ \bibinfo {pages} {1800498} (\bibinfo {year}
  {2019})}\BibitemShut {NoStop}%
\bibitem [{\citenamefont {Valentini}\ \emph {et~al.}(2022)\citenamefont
  {Valentini}, \citenamefont {Borovkov}, \citenamefont {Prada}, \citenamefont
  {Mart\'i-Sánchez}, \citenamefont {Botifoll}, \citenamefont {Hofmann},
  \citenamefont {Arbiol}, \citenamefont {Aguado}, \citenamefont {San-Jos\'e},\
  and\ \citenamefont {Katsaros}}]{valentini2022majorana}%
  \BibitemOpen
  \bibfield  {author} {\bibinfo {author} {\bibfnamefont {M.}~\bibnamefont
  {Valentini}}, \bibinfo {author} {\bibfnamefont {M.}~\bibnamefont {Borovkov}},
  \bibinfo {author} {\bibfnamefont {E.}~\bibnamefont {Prada}}, \bibinfo
  {author} {\bibfnamefont {S.}~\bibnamefont {Mart\'i-Sánchez}}, \bibinfo
  {author} {\bibfnamefont {M.}~\bibnamefont {Botifoll}}, \bibinfo {author}
  {\bibfnamefont {A.}~\bibnamefont {Hofmann}}, \bibinfo {author} {\bibfnamefont
  {J.}~\bibnamefont {Arbiol}}, \bibinfo {author} {\bibfnamefont
  {R.}~\bibnamefont {Aguado}}, \bibinfo {author} {\bibfnamefont
  {P.}~\bibnamefont {San-Jos\'e}}, \ and\ \bibinfo {author} {\bibfnamefont
  {G.}~\bibnamefont {Katsaros}},\ }\href
  {https://www.nature.com/articles/s41586-022-05382-w} {\bibfield  {journal}
  {\bibinfo  {journal} {Nature}\ }\textbf {\bibinfo {volume} {612}},\ \bibinfo
  {pages} {442} (\bibinfo {year} {2022})}\BibitemShut {NoStop}%
\bibitem [{\citenamefont {Hsu}\ \emph {et~al.}(2016)\citenamefont {Hsu},
  \citenamefont {Zhen}, \citenamefont {Stone}, \citenamefont {Joannopoulos},\
  and\ \citenamefont {Solja{\v{c}}i{\'c}}}]{hsu2016bound}%
  \BibitemOpen
  \bibfield  {author} {\bibinfo {author} {\bibfnamefont {C.~W.}\ \bibnamefont
  {Hsu}}, \bibinfo {author} {\bibfnamefont {B.}~\bibnamefont {Zhen}}, \bibinfo
  {author} {\bibfnamefont {A.~D.}\ \bibnamefont {Stone}}, \bibinfo {author}
  {\bibfnamefont {J.~D.}\ \bibnamefont {Joannopoulos}}, \ and\ \bibinfo
  {author} {\bibfnamefont {M.}~\bibnamefont {Solja{\v{c}}i{\'c}}},\ }\href
  {https://www.nature.com/articles/natrevmats201648} {\bibfield  {journal}
  {\bibinfo  {journal} {Nat. Rev. Mater.}\ }\textbf {\bibinfo {volume} {1}},\
  \bibinfo {pages} {1} (\bibinfo {year} {2016})}\BibitemShut {NoStop}%
\bibitem [{\citenamefont {von Neumann}\ and\ \citenamefont
  {Wigner}(1929)}]{vonNeumann-Wigner}%
  \BibitemOpen
  \bibfield  {author} {\bibinfo {author} {\bibfnamefont {J.}~\bibnamefont {von
  Neumann}}\ and\ \bibinfo {author} {\bibfnamefont {E.~P.}\ \bibnamefont
  {Wigner}},\ }\href
  {https://link.springer.com/chapter/10.1007%2F978-3-662-02781-3_19} {\bibfield
   {journal} {\bibinfo  {journal} {Z. Phys.}\ }\textbf {\bibinfo {volume}
  {30}},\ \bibinfo {pages} {465} (\bibinfo {year} {1929})}\BibitemShut
  {NoStop}%
\bibitem [{\citenamefont {Ramos}\ and\ \citenamefont
  {Orellana}(2014)}]{ramos2014bound}%
  \BibitemOpen
  \bibfield  {author} {\bibinfo {author} {\bibfnamefont {J.~P.}\ \bibnamefont
  {Ramos}}\ and\ \bibinfo {author} {\bibfnamefont {P.~A.}\ \bibnamefont
  {Orellana}},\ }\href
  {https://www.sciencedirect.com/science/article/pii/S0921452614005900}
  {\bibfield  {journal} {\bibinfo  {journal} {Phys. B: Condens. Matter}\
  }\textbf {\bibinfo {volume} {455}},\ \bibinfo {pages} {66} (\bibinfo {year}
  {2014})}\BibitemShut {NoStop}%
\bibitem [{\citenamefont {Grez}\ \emph {et~al.}(2022)\citenamefont {Grez},
  \citenamefont {Ramos-Andrade}, \citenamefont {Juri{\v{c}}i{\'c}},\ and\
  \citenamefont {Orellana}}]{grez2022bound}%
  \BibitemOpen
  \bibfield  {author} {\bibinfo {author} {\bibfnamefont {B.}~\bibnamefont
  {Grez}}, \bibinfo {author} {\bibfnamefont {J.}~\bibnamefont {Ramos-Andrade}},
  \bibinfo {author} {\bibfnamefont {V.}~\bibnamefont {Juri{\v{c}}i{\'c}}}, \
  and\ \bibinfo {author} {\bibfnamefont {P.}~\bibnamefont {Orellana}},\ }\href
  {https://journals.aps.org/pra/abstract/10.1103/PhysRevA.106.013719}
  {\bibfield  {journal} {\bibinfo  {journal} {Phys. Rev. A}\ }\textbf {\bibinfo
  {volume} {106}},\ \bibinfo {pages} {013719} (\bibinfo {year}
  {2022})}\BibitemShut {NoStop}%
\bibitem [{\citenamefont {Van~der Wiel}\ \emph {et~al.}(2002)\citenamefont
  {Van~der Wiel}, \citenamefont {De~Franceschi}, \citenamefont {Elzerman},
  \citenamefont {Fujisawa}, \citenamefont {Tarucha},\ and\ \citenamefont
  {Kouwenhoven}}]{van2002electron}%
  \BibitemOpen
  \bibfield  {author} {\bibinfo {author} {\bibfnamefont {W.~G.}\ \bibnamefont
  {Van~der Wiel}}, \bibinfo {author} {\bibfnamefont {S.}~\bibnamefont
  {De~Franceschi}}, \bibinfo {author} {\bibfnamefont {J.~M.}\ \bibnamefont
  {Elzerman}}, \bibinfo {author} {\bibfnamefont {T.}~\bibnamefont {Fujisawa}},
  \bibinfo {author} {\bibfnamefont {S.}~\bibnamefont {Tarucha}}, \ and\
  \bibinfo {author} {\bibfnamefont {L.~P.}\ \bibnamefont {Kouwenhoven}},\
  }\href {https://journals.aps.org/rmp/abstract/10.1103/RevModPhys.75.1}
  {\bibfield  {journal} {\bibinfo  {journal} {Rev. Mod. Phys.}\ }\textbf
  {\bibinfo {volume} {75}},\ \bibinfo {pages} {1} (\bibinfo {year}
  {2002})}\BibitemShut {NoStop}%
\bibitem [{\citenamefont {Hanson}\ \emph {et~al.}(2007)\citenamefont {Hanson},
  \citenamefont {Kouwenhoven}, \citenamefont {Petta}, \citenamefont {Tarucha},\
  and\ \citenamefont {Vandersypen}}]{hanson2007spins}%
  \BibitemOpen
  \bibfield  {author} {\bibinfo {author} {\bibfnamefont {R.}~\bibnamefont
  {Hanson}}, \bibinfo {author} {\bibfnamefont {L.~P.}\ \bibnamefont
  {Kouwenhoven}}, \bibinfo {author} {\bibfnamefont {J.~R.}\ \bibnamefont
  {Petta}}, \bibinfo {author} {\bibfnamefont {S.}~\bibnamefont {Tarucha}}, \
  and\ \bibinfo {author} {\bibfnamefont {L.~M.}\ \bibnamefont {Vandersypen}},\
  }\href {https://journals.aps.org/rmp/abstract/10.1103/RevModPhys.79.1217}
  {\bibfield  {journal} {\bibinfo  {journal} {Rev. Mod. Phys.}\ }\textbf
  {\bibinfo {volume} {79}},\ \bibinfo {pages} {1217} (\bibinfo {year}
  {2007})}\BibitemShut {NoStop}%
\bibitem [{\citenamefont {Holleitner}\ \emph {et~al.}(2001)\citenamefont
  {Holleitner}, \citenamefont {Decker}, \citenamefont {Qin}, \citenamefont
  {Eberl},\ and\ \citenamefont {Blick}}]{holleitner2001coherent}%
  \BibitemOpen
  \bibfield  {author} {\bibinfo {author} {\bibfnamefont {A.}~\bibnamefont
  {Holleitner}}, \bibinfo {author} {\bibfnamefont {C.}~\bibnamefont {Decker}},
  \bibinfo {author} {\bibfnamefont {H.}~\bibnamefont {Qin}}, \bibinfo {author}
  {\bibfnamefont {K.}~\bibnamefont {Eberl}}, \ and\ \bibinfo {author}
  {\bibfnamefont {R.}~\bibnamefont {Blick}},\ }\href
  {https://journals.aps.org/prl/abstract/10.1103/PhysRevLett.87.256802}
  {\bibfield  {journal} {\bibinfo  {journal} {Phys. Rev. Lett.}\ }\textbf
  {\bibinfo {volume} {87}},\ \bibinfo {pages} {256802} (\bibinfo {year}
  {2001})}\BibitemShut {NoStop}%
\bibitem [{\citenamefont {Holleitner}\ \emph {et~al.}(2002)\citenamefont
  {Holleitner}, \citenamefont {Blick}, \citenamefont {Huttel}, \citenamefont
  {Eberl},\ and\ \citenamefont {Kotthaus}}]{holleitner2002probing}%
  \BibitemOpen
  \bibfield  {author} {\bibinfo {author} {\bibfnamefont {A.~W.}\ \bibnamefont
  {Holleitner}}, \bibinfo {author} {\bibfnamefont {R.~H.}\ \bibnamefont
  {Blick}}, \bibinfo {author} {\bibfnamefont {A.~K.}\ \bibnamefont {Huttel}},
  \bibinfo {author} {\bibfnamefont {K.}~\bibnamefont {Eberl}}, \ and\ \bibinfo
  {author} {\bibfnamefont {J.~P.}\ \bibnamefont {Kotthaus}},\ }\href
  {https://www.science.org/doi/full/10.1126/science.1071215} {\bibfield
  {journal} {\bibinfo  {journal} {Science}\ }\textbf {\bibinfo {volume}
  {297}},\ \bibinfo {pages} {70} (\bibinfo {year} {2002})}\BibitemShut
  {NoStop}%
\bibitem [{\citenamefont {Shangguan}\ \emph {et~al.}(2001)\citenamefont
  {Shangguan}, \citenamefont {Yeung}, \citenamefont {Yu},\ and\ \citenamefont
  {Kam}}]{shangguan2001quantum}%
  \BibitemOpen
  \bibfield  {author} {\bibinfo {author} {\bibfnamefont {W.~Z.}\ \bibnamefont
  {Shangguan}}, \bibinfo {author} {\bibfnamefont {T.~C.~A.}\ \bibnamefont
  {Yeung}}, \bibinfo {author} {\bibfnamefont {Y.~B.}\ \bibnamefont {Yu}}, \
  and\ \bibinfo {author} {\bibfnamefont {C.~H.}\ \bibnamefont {Kam}},\ }\href
  {https://journals.aps.org/prb/abstract/10.1103/PhysRevB.63.235323} {\bibfield
   {journal} {\bibinfo  {journal} {Phys. Rev. B}\ }\textbf {\bibinfo {volume}
  {63}},\ \bibinfo {pages} {235323} (\bibinfo {year} {2001})}\BibitemShut
  {NoStop}%
\bibitem [{\citenamefont {Orellana}\ \emph {et~al.}(2003)\citenamefont
  {Orellana}, \citenamefont {Dominguez-Adame}, \citenamefont {G{\'o}mez},\ and\
  \citenamefont {De~Guevara}}]{orellana2003transport}%
  \BibitemOpen
  \bibfield  {author} {\bibinfo {author} {\bibfnamefont {P.~A.}\ \bibnamefont
  {Orellana}}, \bibinfo {author} {\bibfnamefont {F.}~\bibnamefont
  {Dominguez-Adame}}, \bibinfo {author} {\bibfnamefont {I.}~\bibnamefont
  {G{\'o}mez}}, \ and\ \bibinfo {author} {\bibfnamefont {M.~L.~L.}\
  \bibnamefont {De~Guevara}},\ }\href
  {https://journals.aps.org/prb/abstract/10.1103/PhysRevB.67.085321} {\bibfield
   {journal} {\bibinfo  {journal} {Phys. Rev. B}\ }\textbf {\bibinfo {volume}
  {67}},\ \bibinfo {pages} {085321} (\bibinfo {year} {2003})}\BibitemShut
  {NoStop}%
\bibitem [{\citenamefont {Hofstetter}\ \emph {et~al.}(2001)\citenamefont
  {Hofstetter}, \citenamefont {K{\"o}nig},\ and\ \citenamefont
  {Schoeller}}]{hofstetter2001kondo}%
  \BibitemOpen
  \bibfield  {author} {\bibinfo {author} {\bibfnamefont {W.}~\bibnamefont
  {Hofstetter}}, \bibinfo {author} {\bibfnamefont {J.}~\bibnamefont
  {K{\"o}nig}}, \ and\ \bibinfo {author} {\bibfnamefont {H.}~\bibnamefont
  {Schoeller}},\ }\href
  {https://journals.aps.org/prl/abstract/10.1103/PhysRevLett.87.156803}
  {\bibfield  {journal} {\bibinfo  {journal} {Phys. Rev. Lett.}\ }\textbf
  {\bibinfo {volume} {87}},\ \bibinfo {pages} {156803} (\bibinfo {year}
  {2001})}\BibitemShut {NoStop}%
\bibitem [{\citenamefont {G{\'o}rski}\ and\ \citenamefont
  {Kucab}(2019)}]{gorski2019spin}%
  \BibitemOpen
  \bibfield  {author} {\bibinfo {author} {\bibfnamefont {G.}~\bibnamefont
  {G{\'o}rski}}\ and\ \bibinfo {author} {\bibfnamefont {K.}~\bibnamefont
  {Kucab}},\ }\href
  {https://onlinelibrary.wiley.com/doi/full/10.1002/pssb.201800492} {\bibfield
  {journal} {\bibinfo  {journal} {Phys. Status Solidi B}\ }\textbf {\bibinfo
  {volume} {256}},\ \bibinfo {pages} {1800492} (\bibinfo {year}
  {2019})}\BibitemShut {NoStop}%
\bibitem [{\citenamefont {Chi}\ \emph {et~al.}(2007)\citenamefont {Chi},
  \citenamefont {Liu},\ and\ \citenamefont {Sun}}]{chi2007fano}%
  \BibitemOpen
  \bibfield  {author} {\bibinfo {author} {\bibfnamefont {F.}~\bibnamefont
  {Chi}}, \bibinfo {author} {\bibfnamefont {J.-L.}\ \bibnamefont {Liu}}, \ and\
  \bibinfo {author} {\bibfnamefont {L.-L.}\ \bibnamefont {Sun}},\ }\href
  {https://aip.scitation.org/doi/full/10.1063/1.2720097?casa_token=9YYOouWNqmYAAAAA%3AGCsYLBnrx5UbnsCUghrPc9VyH921L3V3db7WEnURHt8KMSyZxID6vALJqvB6NL_vw5fXqlWv5wBH}
  {\bibfield  {journal} {\bibinfo  {journal} {J. Appl. Phys}\ }\textbf
  {\bibinfo {volume} {101}},\ \bibinfo {pages} {093704} (\bibinfo {year}
  {2007})}\BibitemShut {NoStop}%
\bibitem [{\citenamefont {Kubala}\ and\ \citenamefont
  {K{\"o}nig}(2002)}]{kubala2002flux}%
  \BibitemOpen
  \bibfield  {author} {\bibinfo {author} {\bibfnamefont {B.}~\bibnamefont
  {Kubala}}\ and\ \bibinfo {author} {\bibfnamefont {J.}~\bibnamefont
  {K{\"o}nig}},\ }\href
  {https://journals.aps.org/prb/abstract/10.1103/PhysRevB.65.245301} {\bibfield
   {journal} {\bibinfo  {journal} {Phys. Rev. B}\ }\textbf {\bibinfo {volume}
  {65}},\ \bibinfo {pages} {245301} (\bibinfo {year} {2002})}\BibitemShut
  {NoStop}%
\bibitem [{\citenamefont {Fano}(1961)}]{fano1961effects}%
  \BibitemOpen
  \bibfield  {author} {\bibinfo {author} {\bibfnamefont {U.}~\bibnamefont
  {Fano}},\ }\href
  {https://journals.aps.org/pr/abstract/10.1103/PhysRev.124.1866} {\bibfield
  {journal} {\bibinfo  {journal} {Phys. Rev.}\ }\textbf {\bibinfo {volume}
  {124}},\ \bibinfo {pages} {1866} (\bibinfo {year} {1961})}\BibitemShut
  {NoStop}%
\bibitem [{\citenamefont {Miroshnichenko}\ \emph {et~al.}(2010)\citenamefont
  {Miroshnichenko}, \citenamefont {Flach},\ and\ \citenamefont
  {Kivshar}}]{miroshnichenko2010fano}%
  \BibitemOpen
  \bibfield  {author} {\bibinfo {author} {\bibfnamefont {A.~E.}\ \bibnamefont
  {Miroshnichenko}}, \bibinfo {author} {\bibfnamefont {S.}~\bibnamefont
  {Flach}}, \ and\ \bibinfo {author} {\bibfnamefont {Y.~S.}\ \bibnamefont
  {Kivshar}},\ }\href
  {https://journals.aps.org/rmp/abstract/10.1103/RevModPhys.82.2257} {\bibfield
   {journal} {\bibinfo  {journal} {Rev. Mod. Phys.}\ }\textbf {\bibinfo
  {volume} {82}},\ \bibinfo {pages} {2257} (\bibinfo {year}
  {2010})}\BibitemShut {NoStop}%
\bibitem [{\citenamefont {Ladr\'on~de Guevara}\ \emph
  {et~al.}(2003)\citenamefont {Ladr\'on~de Guevara}, \citenamefont {Claro},\
  and\ \citenamefont {Orellana}}]{de2003ghost}%
  \BibitemOpen
  \bibfield  {author} {\bibinfo {author} {\bibfnamefont {M.~L.}\ \bibnamefont
  {Ladr\'on~de Guevara}}, \bibinfo {author} {\bibfnamefont {F.}~\bibnamefont
  {Claro}}, \ and\ \bibinfo {author} {\bibfnamefont {P.~A.}\ \bibnamefont
  {Orellana}},\ }\href
  {https://journals.aps.org/prb/abstract/10.1103/PhysRevB.67.195335} {\bibfield
   {journal} {\bibinfo  {journal} {Phys. Rev. B}\ }\textbf {\bibinfo {volume}
  {67}},\ \bibinfo {pages} {195335} (\bibinfo {year} {2003})}\BibitemShut
  {NoStop}%
\bibitem [{\citenamefont {Gong}\ \emph {et~al.}(2014)\citenamefont {Gong},
  \citenamefont {Zhang}, \citenamefont {Li}, \citenamefont {Yi},\ and\
  \citenamefont {Zheng}}]{gong2014detection}%
  \BibitemOpen
  \bibfield  {author} {\bibinfo {author} {\bibfnamefont {W.-J.}\ \bibnamefont
  {Gong}}, \bibinfo {author} {\bibfnamefont {S.-F.}\ \bibnamefont {Zhang}},
  \bibinfo {author} {\bibfnamefont {Z.-C.}\ \bibnamefont {Li}}, \bibinfo
  {author} {\bibfnamefont {G.}~\bibnamefont {Yi}}, \ and\ \bibinfo {author}
  {\bibfnamefont {Y.-S.}\ \bibnamefont {Zheng}},\ }\href
  {https://journals.aps.org/prb/abstract/10.1103/PhysRevB.89.245413} {\bibfield
   {journal} {\bibinfo  {journal} {Phys. Rev. B}\ }\textbf {\bibinfo {volume}
  {89}},\ \bibinfo {pages} {245413} (\bibinfo {year} {2014})}\BibitemShut
  {NoStop}%
\bibitem [{\citenamefont {Chi}\ \emph {et~al.}(2021)\citenamefont {Chi},
  \citenamefont {Wang}, \citenamefont {He}, \citenamefont {Fu}, \citenamefont
  {Zhang}, \citenamefont {Zhang}, \citenamefont {Wang},\ and\ \citenamefont
  {Lu}}]{chi2021quantum}%
  \BibitemOpen
  \bibfield  {author} {\bibinfo {author} {\bibfnamefont {F.}~\bibnamefont
  {Chi}}, \bibinfo {author} {\bibfnamefont {J.}~\bibnamefont {Wang}}, \bibinfo
  {author} {\bibfnamefont {T.-Y.}\ \bibnamefont {He}}, \bibinfo {author}
  {\bibfnamefont {Z.-G.}\ \bibnamefont {Fu}}, \bibinfo {author} {\bibfnamefont
  {P.}~\bibnamefont {Zhang}}, \bibinfo {author} {\bibfnamefont {X.-W.}\
  \bibnamefont {Zhang}}, \bibinfo {author} {\bibfnamefont {L.}~\bibnamefont
  {Wang}}, \ and\ \bibinfo {author} {\bibfnamefont {Z.}~\bibnamefont {Lu}},\
  }\href {https://www.frontiersin.org/articles/10.3389/fphy.2020.631031/full}
  {\bibfield  {journal} {\bibinfo  {journal} {Front. Phys.}\ }\textbf {\bibinfo
  {volume} {8}},\ \bibinfo {pages} {631031} (\bibinfo {year}
  {2021})}\BibitemShut {NoStop}%
\bibitem [{\citenamefont {Liu}\ and\ \citenamefont
  {Baranger}(2011)}]{liu2011detecting}%
  \BibitemOpen
  \bibfield  {author} {\bibinfo {author} {\bibfnamefont {D.~E.}\ \bibnamefont
  {Liu}}\ and\ \bibinfo {author} {\bibfnamefont {H.~U.}\ \bibnamefont
  {Baranger}},\ }\href
  {https://journals.aps.org/prb/abstract/10.1103/PhysRevB.84.201308} {\bibfield
   {journal} {\bibinfo  {journal} {Phys. Rev. B}\ }\textbf {\bibinfo {volume}
  {84}},\ \bibinfo {pages} {201308} (\bibinfo {year} {2011})}\BibitemShut
  {NoStop}%
\bibitem [{\citenamefont {Vernek}\ \emph {et~al.}(2014)\citenamefont {Vernek},
  \citenamefont {Penteado}, \citenamefont {Seridonio},\ and\ \citenamefont
  {Egues}}]{vernek2014subtle}%
  \BibitemOpen
  \bibfield  {author} {\bibinfo {author} {\bibfnamefont {E.}~\bibnamefont
  {Vernek}}, \bibinfo {author} {\bibfnamefont {P.~H.}\ \bibnamefont
  {Penteado}}, \bibinfo {author} {\bibfnamefont {A.~C.}\ \bibnamefont
  {Seridonio}}, \ and\ \bibinfo {author} {\bibfnamefont {J.~C.}\ \bibnamefont
  {Egues}},\ }\href
  {https://journals.aps.org/prb/abstract/10.1103/PhysRevB.89.165314} {\bibfield
   {journal} {\bibinfo  {journal} {Phys. Rev. B}\ }\textbf {\bibinfo {volume}
  {89}},\ \bibinfo {pages} {165314} (\bibinfo {year} {2014})}\BibitemShut
  {NoStop}%
\bibitem [{\citenamefont {Deng}\ \emph {et~al.}(2016)\citenamefont {Deng},
  \citenamefont {Vaitiek{\.e}nas}, \citenamefont {Hansen}, \citenamefont
  {Danon}, \citenamefont {Leijnse}, \citenamefont {Flensberg}, \citenamefont
  {Nyg{\aa}rd}, \citenamefont {Krogstrup},\ and\ \citenamefont
  {Marcus}}]{deng2016majorana}%
  \BibitemOpen
  \bibfield  {author} {\bibinfo {author} {\bibfnamefont {M.}~\bibnamefont
  {Deng}}, \bibinfo {author} {\bibfnamefont {S.}~\bibnamefont
  {Vaitiek{\.e}nas}}, \bibinfo {author} {\bibfnamefont {E.~B.}\ \bibnamefont
  {Hansen}}, \bibinfo {author} {\bibfnamefont {J.}~\bibnamefont {Danon}},
  \bibinfo {author} {\bibfnamefont {M.}~\bibnamefont {Leijnse}}, \bibinfo
  {author} {\bibfnamefont {K.}~\bibnamefont {Flensberg}}, \bibinfo {author}
  {\bibfnamefont {J.}~\bibnamefont {Nyg{\aa}rd}}, \bibinfo {author}
  {\bibfnamefont {P.}~\bibnamefont {Krogstrup}}, \ and\ \bibinfo {author}
  {\bibfnamefont {C.~M.}\ \bibnamefont {Marcus}},\ }\href
  {https://www.science.org/doi/full/10.1126/science.aaf3961} {\bibfield
  {journal} {\bibinfo  {journal} {Science}\ }\textbf {\bibinfo {volume}
  {354}},\ \bibinfo {pages} {1557} (\bibinfo {year} {2016})}\BibitemShut
  {NoStop}%
\bibitem [{\citenamefont {Ricco}\ \emph {et~al.}(2016)\citenamefont {Ricco},
  \citenamefont {Marques}, \citenamefont {Dessotti}, \citenamefont {Machado},
  \citenamefont {De~Souza},\ and\ \citenamefont {Seridonio}}]{ricco2016decay}%
  \BibitemOpen
  \bibfield  {author} {\bibinfo {author} {\bibfnamefont {L.~S.}\ \bibnamefont
  {Ricco}}, \bibinfo {author} {\bibfnamefont {Y.}~\bibnamefont {Marques}},
  \bibinfo {author} {\bibfnamefont {F.~A.}\ \bibnamefont {Dessotti}}, \bibinfo
  {author} {\bibfnamefont {R.~S.}\ \bibnamefont {Machado}}, \bibinfo {author}
  {\bibfnamefont {M.}~\bibnamefont {De~Souza}}, \ and\ \bibinfo {author}
  {\bibfnamefont {A.~C.}\ \bibnamefont {Seridonio}},\ }\href
  {https://journals.aps.org/prb/abstract/10.1103/PhysRevB.93.165116} {\bibfield
   {journal} {\bibinfo  {journal} {Phys. Rev. B}\ }\textbf {\bibinfo {volume}
  {93}},\ \bibinfo {pages} {165116} (\bibinfo {year} {2016})}\BibitemShut
  {NoStop}%
\bibitem [{\citenamefont {Guessi}\ \emph {et~al.}(2017)\citenamefont {Guessi},
  \citenamefont {Dessotti}, \citenamefont {Marques}, \citenamefont {Ricco},
  \citenamefont {Pereira}, \citenamefont {Menegasso}, \citenamefont
  {De~Souza},\ and\ \citenamefont {Seridonio}}]{guessi2017encrypting}%
  \BibitemOpen
  \bibfield  {author} {\bibinfo {author} {\bibfnamefont {L.~H.}\ \bibnamefont
  {Guessi}}, \bibinfo {author} {\bibfnamefont {F.~A.}\ \bibnamefont
  {Dessotti}}, \bibinfo {author} {\bibfnamefont {Y.}~\bibnamefont {Marques}},
  \bibinfo {author} {\bibfnamefont {L.~S.}\ \bibnamefont {Ricco}}, \bibinfo
  {author} {\bibfnamefont {G.~M.}\ \bibnamefont {Pereira}}, \bibinfo {author}
  {\bibfnamefont {P.}~\bibnamefont {Menegasso}}, \bibinfo {author}
  {\bibfnamefont {M.}~\bibnamefont {De~Souza}}, \ and\ \bibinfo {author}
  {\bibfnamefont {A.~C.}\ \bibnamefont {Seridonio}},\ }\href
  {https://journals.aps.org/prb/abstract/10.1103/PhysRevB.96.041114} {\bibfield
   {journal} {\bibinfo  {journal} {Phys. Rev. B}\ }\textbf {\bibinfo {volume}
  {96}},\ \bibinfo {pages} {041114(R)} (\bibinfo {year} {2017})}\BibitemShut
  {NoStop}%
\bibitem [{\citenamefont {Ruiz-Tijerina}\ \emph {et~al.}(2015)\citenamefont
  {Ruiz-Tijerina}, \citenamefont {Vernek}, \citenamefont {Dias~da Silva},\ and\
  \citenamefont {Egues}}]{ruiz2015interaction}%
  \BibitemOpen
  \bibfield  {author} {\bibinfo {author} {\bibfnamefont {D.~A.}\ \bibnamefont
  {Ruiz-Tijerina}}, \bibinfo {author} {\bibfnamefont {E.}~\bibnamefont
  {Vernek}}, \bibinfo {author} {\bibfnamefont {L.~G. G.~V.}\ \bibnamefont
  {Dias~da Silva}}, \ and\ \bibinfo {author} {\bibfnamefont {J.~C.}\
  \bibnamefont {Egues}},\ }\href
  {https://journals.aps.org/prb/abstract/10.1103/PhysRevB.91.115435} {\bibfield
   {journal} {\bibinfo  {journal} {Phys. Rev. B}\ }\textbf {\bibinfo {volume}
  {91}},\ \bibinfo {pages} {115435} (\bibinfo {year} {2015})}\BibitemShut
  {NoStop}%
\bibitem [{\citenamefont {Datta}(2005)}]{datta2005quantum}%
  \BibitemOpen
  \bibfield  {author} {\bibinfo {author} {\bibfnamefont {S.}~\bibnamefont
  {Datta}},\ }\href
  {https://www.cambridge.org/cl/academic/subjects/engineering/electronic-optoelectronic-devices-and-nanotechnology/quantum-transport-atom-transistor?format=HB&isbn=9780521631457}
  {\emph {\bibinfo {title} {Quantum transport: atom to transistor}}}\ (\bibinfo
   {publisher} {Cambridge university press},\ \bibinfo {year}
  {2005})\BibitemShut {NoStop}%
\bibitem [{\citenamefont {Meir}\ and\ \citenamefont
  {Wingreen}(1992)}]{meir1992landauer}%
  \BibitemOpen
  \bibfield  {author} {\bibinfo {author} {\bibfnamefont {Y.}~\bibnamefont
  {Meir}}\ and\ \bibinfo {author} {\bibfnamefont {N.~S.}\ \bibnamefont
  {Wingreen}},\ }\href
  {https://journals.aps.org/prl/abstract/10.1103/PhysRevLett.68.2512}
  {\bibfield  {journal} {\bibinfo  {journal} {Phys. Rev. Lett.}\ }\textbf
  {\bibinfo {volume} {68}},\ \bibinfo {pages} {2512} (\bibinfo {year}
  {1992})}\BibitemShut {NoStop}%
\bibitem [{\citenamefont {Calle}\ \emph {et~al.}(2020)\citenamefont {Calle},
  \citenamefont {Pacheco}, \citenamefont {Orellana},\ and\ \citenamefont
  {Ot{\'a}lora}}]{calle2020fano}%
  \BibitemOpen
  \bibfield  {author} {\bibinfo {author} {\bibfnamefont {A.~M.}\ \bibnamefont
  {Calle}}, \bibinfo {author} {\bibfnamefont {M.}~\bibnamefont {Pacheco}},
  \bibinfo {author} {\bibfnamefont {P.~A.}\ \bibnamefont {Orellana}}, \ and\
  \bibinfo {author} {\bibfnamefont {J.~A.}\ \bibnamefont {Ot{\'a}lora}},\
  }\href {https://onlinelibrary.wiley.com/doi/abs/10.1002/andp.201900409}
  {\bibfield  {journal} {\bibinfo  {journal} {Ann. Phys. (Berlin)}\ }\textbf
  {\bibinfo {volume} {532}},\ \bibinfo {pages} {1900409} (\bibinfo {year}
  {2020})}\BibitemShut {NoStop}%
\end{thebibliography}%

\end{document}